\global\def\draftcontrol{0}

   \def\versionno{ alpha prime charged BH -- draft   }
\catcode`\@=11

\expandafter\ifx\csname draftcontrol\endcsname\relax\global\def\draftcontrol{0}
\fi

{\count255=\time\divide\count255 by 60
\xdef\hourmin{\number\count255}
\multiply\count255 by-60\advance\count255 by\time
\xdef\hourmin{\hourmin:\ifnum\count255<10 0\fi\the\count255}}
\def\draftdate{\number\month/\number\day/\number\year\ \ \ \hourmin }

\newcommand\makepapertitle{\par
  \begingroup
    \renewcommand\thefootnote{\@fnsymbol\c@footnote}%
    \def\@makefnmark{\rlap{\@textsuperscript{\normalfont\@thefnmark}}}%
    \long\def\@makefntext##1{\parindent 1em\noindent
            \hb@xt@1.8em{%
                \hss\@textsuperscript{\normalfont\@thefnmark}}##1}%
     \newpage
     \global\@topnum\z@   % Prevents figures from going at top of page.
     \@makepapertitle
     \thispagestyle{empty}\@thanks
  \endgroup
  \setcounter{footnote}{0}%
  \global\let\thanks\relax
  \global\let\makepapertitle\relax
  \global\let\@makepapertitle\relax
  \global\let\@thanks\@empty
  \global\let\@author\@empty
  \global\let\@date\@empty
  \global\let\@title\@empty
  \global\let\title\relax
  \global\let\author\relax
  \global\let\date\relax
  \global\let\and\relax
  \def\version{\let\version\@version\@gobble}
}
\def\@makepapertitle{%
  \newpage
   \ifnum\draftcontrol=1 {}
   \version\versionno
   \vskip 3em%
   \else
   \hfill\hbox to 3cm {\parbox{4cm}{\@pubnum}\hss}%
   \vskip 3em%
   \fi
   \begin{center}%
   \let \footnote \thanks
     {\LARGE {\@title}}%
     \vskip 1.5em%
     {\normalsize%\large
       \lineskip .5em%
       \begin{tabular}[t]{c}%
         \@author
       \end{tabular}\par}%
     \vskip 1.5em%
     {\@bstract}%
     \end{center}%
     \vskip 1.5em
     \@date%
   \par
}

\gdef\@pubnum{}
\def\pubnum#1{%
  \gdef\@pubnum{#1}}

\gdef\@bstract{}
\def\Abstract#1{%
  \gdef\@bstract{%
   \parbox{\textwidth-0pc}{%
   \centerline{\bf Abstract}\penalty1000%
\kern.2cm%
\noindent%\abstractfont \baselineskip=12pt
\renewcommand\baselinestretch{1.0}%
{#1}}}
}

\def\ps@paper{\let\@mkboth\@gobbletwo%
     \ifnum\draftcontrol=1
	\def\@oddfoot{\hbox to \textwidth{\tiny \versionno \hfil\tiny\draftdate}%
	\hskip -\textwidth \hbox to \textwidth{\hfil\rm\thepage\hfil}}%
     \else\def\@oddfoot{\hbox to \textwidth{\hfil\rm\thepage\hfil}}
     \fi
     \let\@evenfoot\@oddfoot
}
\def\body{\clearpage
          \pagestyle{paper}
	}
\def\@version#1{\ifnum\draftcontrol=1
\typeout{}\typeout{#1}\typeout{}
\vskip3mm\centerline{\hbox{\fbox{\normalsize{\tt DRAFT -- #1 -- }
                   {\draftdate}}}}\vskip3mm
\fi}
\let\version\@version
\long\def\eqlabel#1{\ifnum\draftcontrol=1
                    \tag@false  % there are some problems with multline without this
                    \tag*{(\theequation) \hbox to -0.2cm{\hspace{0cm}\small{#1}\hss}}
                    \refstepcounter{equation}
                    \edef\@currentlabel{\theequation}
                    \ltx@label{#1}          % use old LaTeX \label instead of new definition
                    \else
                    \label{#1}
                    \fi
                    }
\let\st@bibitem\@bibitem
\let\st@lbibitem\@lbibitem
\ifnum\draftcontrol=1
  \def\@bibitem#1{%
    \st@bibitem{#1}\a@@label{#1}\ignorespaces}
  \def\@lbibitem[#1]#2{%
    \st@lbibitem[#1]{#2}\a@@label{#2}\ignorespaces}
  \def\a@@label#1{%
    \gdef\a@lab{\smash{\normalfont\small#1}}
    \ifvmode
      \if@inlabel
        \global\setbox\@labels\hbox{%
          \llap{\a@lab\let\a@lab\relax
                \kern\@totalleftmargin\kern\marginparsep}%
          \box\@labels}%
      \fi
    \fi}
\fi
\documentclass[12pt,letterpaper]{article}
\usepackage{amsmath,amssymb,array,calc,rotating,epsfig,psfrag}
\usepackage[nosort]{cite}
\ifnum\draftcontrol=1
\fi
\renewcommand\baselinestretch{1.25}
\renewcommand\section{\@startsection {section}{1}{\z@}%
                                   {-3.5ex \@plus -1ex \@minus -.2ex}%
                                   {2.3ex \@plus.2ex}%
                                   {\normalfont\large\bfseries}}
\renewcommand\subsection{\@startsection{subsection}{2}{\z@}%
                                   {-3.25ex\@plus -1ex \@minus -.2ex}%
                                   {1.5ex \@plus .2ex}%
                                   {\normalfont\normalsize\bfseries}}
\renewcommand\subsubsection{\@startsection{subsubsection}{3}{\z@}%
                                   {-3.25ex\@plus -1ex \@minus -.2ex}%
                                   {1.5ex \@plus .2ex}%
                                   {\normalfont\normalsize\it}}
\renewcommand\paragraph{\@startsection{paragraph}{4}{\z@}%
                                   {-3.25ex\@plus -1ex \@minus -.2ex}%
                                   {1.5ex \@plus .2ex}%
                                   {\normalfont\normalsize\bf}}

\numberwithin{equation}{section}

\def\ie{{\it i.e.}}

\def\revise#1       {\raisebox{-0em}{\rule{3pt}{1em}}%
                     \marginpar{\raisebox{.5em}{\vrule width3pt\
                     \vrule width0pt height 0pt depth0.5em
                     \hbox to 0cm{\hspace{0cm}{%
                     \parbox[t]{4em}{\raggedright\footnotesize{#1}}}\hss}}}}

\newcommand\nxt[1]  {\\\fnxt#1}

\def\calb         {{\cal B}}
\def\calc         {{\cal C}}

\def\calf         {{\cal F}}

\def\cali         {{\cal I}}

\def\calm         {{\cal M}}
\def\caln         {{\cal N}}
\def\calo         {{\cal O}}

\def\del          {\partial}

\def\tr           {\mathop{\rm Tr}}

\def\sqr#1#2{{\vcenter{\vbox{\hrule height.#2pt
 \hbox{\vrule width.#2pt height#1pt \kern#1pt
 \vrule width.#2pt}\hrule height.#2pt}}}}
\def\square{%
  \mathop{\mathchoice{\sqr{12}{15}}{\sqr{9}{12}}{\sqr{6.3}{9}}{\sqr{4.5}{9}}}}

\newcommand{\ft}[2]{{\textstyle{\frac{#1}{#2}}}}

\def\a{\alpha}
\def\k{\kappa}
\def\e{\epsilon}
\def\l{\lambda}
\def\tq{\tilde{q}}
\def\rp{r_+}
\def\xp{x_+}
\def\tQ{\tilde{Q}}
\def\g{\gamma}
\def\dd{\delta}
\def\hw{\hat{\Omega}}
\def\he{\hat{E}}
\def\hs{\hat{S}}
\def\hmu{\hat{\mu}_q}
\def\hf{\hat{F}}
\def\r{\rho}

\catcode`\@=12

\begin{document}

%%%
%%%%%% text starts here
%%%%%%%%%

\title{Higher derivative corrections to near-extremal black holes in type IIB supergravity}

\pubnum{%
UWO-TH-06/04
}
\date{April 2006}

\author{
Alex Buchel\\[0.4cm]
\it Department of Applied Mathematics\\
\it University of Western Ontario\\
\it London, Ontario N6A 5B7, Canada\\[0.2cm]
\it Perimeter Institute for Theoretical Physics\\
\it Waterloo, Ontario N2J 2W9, Canada
}

\Abstract{
We discuss string theory $\a'$ corrections to charged near-extremal
black 3-branes/black holes in type IIB supergravity. We find that
supersymmetric global $AdS_5\times S^5$ geometry is not corrected to
leading order in $\a'$, while charged or non-extremal black
3-branes/black hole geometries receive $\a'$ corrections.  Following
gauge theory-string theory correspondence the thermodynamics of these
geometries is mapped to the thermodynamics of large-$n_c$ $\caln=4$
supersymmetric Yang-Mills theory at finite (large) 't Hooft coupling
with the $U(1)_R$-charge chemical potential.  We use holographic
renormalization to compute the Gibbs free energy and the ADM mass of
the near-extremal solutions. The remaining thermodynamic potentials
are evaluated enforcing the first law of thermodynamics.  We present
analytic expressions for the $\a'$ corrected thermodynamics of black
holes in $AdS_5\times S^5$ and the thermodynamics of charged black
3-branes with identical chemical potentials for $[U(1)_R]^3$ charges
and large (compare to chemical potential) temperature. We compute
$\a'$ corrections to Hawking-Page phase transition. We find that for
nonzero chemical potential thermodynamics of near-extremal black
3-brane solution receives $\ln T$ correction to leading order in
$\a'$.
}

%\enlargethispage{1.5cm}

\makepapertitle

\body

\version\versionno

\section{Introduction}
Gauge theory-string theory correspondence of Maldacena \cite{m9711,m2}
presents analytical tools to study non-perturbative aspects of strongly
coupled gauge theories. In particular, by mapping finite temperature
thermodynamics of gauge theories to thermodynamics of black holes in
asymptotically anti-de-Sitter geometries it allows to use field
theoretic intuition to uncover new phenomena in black hole
physics. The most studied example of the gauge theory/string theory correspondence
is that between $\caln=4$ supersymmetric $SU(n_c)$ Yang-Mills (SYM)
and $AdS_5\times S^5$ geometry in type IIB string theory. On the
string theory side the latter duality is the simplest in the 't Hooft
( large $n_c$ ) limit and for large values of the 't Hooft coupling
$\l=n_c g_{YM}^2\gg 1$. In this regime both the string loop
corrections $\propto \ft{1}{n_c}$ and the string theory $\a'$
corrections $\propto \l^{-1/2}$ are suppressed and the supergravity
approximation is reliable. Introducing finite temperature on the
Yang-Mills side corresponds to introducing horizon ( study nonextremal
black hole solutions ) in asymptotic $AdS_5\times S^5$ type IIB
supergravity. On the supergravity side one can study black holes with
flat spatial horizon $\k=0$ ( these are often referred to as black
branes ), or with spatial curvature $\k=\{-1,1\}$.  The first example
is a supergravity dual to a finite temperature $\caln=4$ SYM on a
space-time four-manifold $\calm_4\equiv R^{3,1}$ originally studied in
\cite{gkp}. The other examples correspond to finite temperature
$\caln=4$ SYM on space-time manifolds $\calm_4\equiv R\times AdS_3$
\cite{em,rc1} and $\calm_4\equiv R\times S_3$ correspondingly.  Here, the
existence of the second energy scale in the problem opens up a
possibility for interesting phase transitions.  Indeed, in \cite{w98}
it was explained that the thermal Hawking-Page phase transition for
$\k=1$ AdS black holes is mapped to a finite temperature
confinement/deconfinement phase transition of the $\caln=4$ SYM on
$R\times S^3$.

Supersymmetric $\caln=4$ Yang Mills theory has $SO(6)_R$ symmetry
realized as a global symmetry of the dual supergravity solution.  Thus
one can generalize SYM thermodynamics by introducing chemical
potentials for $[U(1)]^3\subset SO(6)$ charges of the R-symmetry.  On
the gravity side this corresponds to study Riessner-Nordstrom black
holes in $D=5$ $\caln=8$ gauged supergravity (the so-called STU
models) \cite{stu} or rotating black hole solutions in type IIB
supergravity \cite{stulift}. Thermodynamics of such black holes has
been extensively studied in \cite{th1,th2,th3,th4}.

In this paper we study string theory $\a'$ corrections to 
thermodynamics of neutral and charged black holes in 
asymptotic $AdS_5\times S^5$ geometry of  type IIB 
supergravity. 
Previously, only  thermodynamics  of neutral $\k=0$ black hole 
has been analyzed in the literature\footnote{$\a'$-corrected thermodynamics 
of near-extremal $(D1,D3)$ bound states was discussed in \cite{rc2}. 
} \cite{gkt,pt}. Furthermore, 
it was argued in \cite{bb} that consistency of the hydrodynamic 
description\footnote{Shear viscosity of the strongly coupled $\caln=4$ SYM
at finite 't Hooft coupling was originally computed in \cite{bls}.}
of $\caln=4$ SYM at finite 't Hooft coupling 
provides a highly nontrivial consistency check on 
the $\a'$ thermodynamics of the 
$\k=0$ neutral black holes.

The paper is organized as follows. In the next section we present type
IIB supergravity solution representing the ten-dimensional
uplift\footnote{We correct some misprints in \cite{stulift}.} of the
STU models \cite{stu}.  In section 3 we discuss our conventions and
set up computational framework.  For technical reasons we consider only
the cases of neutral nonextremal black holes with spatial curvature
$\k=\{0,1\}$, or black holes with spatial curvature $\k=\{0,1\}$
and identical charges under $U(1)^3\subset SO(6)$ R-symmetry.  In
section 4 we present $\a'$-corrected equations of motion for these
supergravity geometries. We find a convenient gauge in which all these
equations decouple, and can be solved analytically for any values of
the temperature and the chemical potential, apart from the (decoupled)
$S^5$ warp factor second order ordinary differential equation.  The
latter equation we managed to solve analytically only for neutral
$\k=0$ black holes\footnote{This case was previously considered in
\cite{pt}.}, and for charged $\k=0$ black holes in the high
temperature $T$ (compare to the chemical potential $\mu_q$)
approximation, \ie,\ $T\gg \mu_q$. In section 5 we consider
thermodynamics of the $\a'$ corrected nonextremal solutions.  We
discuss relevant aspects of $\a'$ holographic renormalization and
study thermodynamics of the $\k=\{0,1\}$ neutral black holes. Even
though we do not have explicit analytic solution for the $S^5$ warp
factor for nonextremal black holes with spatial curvature
$\k=1$, we managed to compute analytically the Helmholtz free
energy $F$ and the ADM mass $E$ of the black holes. Turns out, the
neutral black hole entropy computed from the thermodynamic relation
$S=(E-F)/T$ automatically satisfied the first law of thermodynamics.
We find that the extremal global $AdS_5\times S^5$ geometry does not
receive $\a'$ corrections to leading order.  We evaluate leading $\a'$
corrections to the Hawking-Page transition temperature, or
equivalently, leading 't Hooft coupling corrections to the finite
temperature confinement-deconfinement phase transition of $\caln=4$
supersymmetric Yang-Mills theory on $S^3$.  Finally, we discuss
thermodynamics of charged black holes.  Here, the knowledge of the
$S^5$ warp factor of the underlying geometry is crucial. We managed
to find analytic solution only for $\k=0$ charged black holes to
leading order in $\ft {\mu_q}{T}\ll 1$. We use holographic
renormalization to compute the Gibbs free energy and the ADM mass of
the black hole, and evaluate remaining thermodynamic properties ( \ie,
the entropy, the physical R-symmetry charges and the conjugate
chemical potentials) enforcing the first law of thermodynamics.  Rather
surprisingly, we find that thermodynamics of charged $\k=0$ black
holes receives to leading order in $\a'$ correction $\propto
\ln \ft {T}{\e}$. We comment on the appearance of an arbitrary energy scale 
$\e$ in the system in the conclusion section.    
Some particularly technical details can be found in Appendix.

\section{Embedding STU models in type IIB SUGRA}
The ten dimensional lift of STU models \cite{stu} was explained in \cite{stulift}.
Of course, such uplifted solutions are fundamental to construction of $\a'$-corrected
black holes discuss in this paper. For this reason we write down explicitly corresponding  
10d supergravity solutions fixing some misprints in \cite{stulift}.  

The ten dimensional metric of a black hole with a spatial horizon curvature 
$\k$, the nonextremality parameter $\mu$, and charges $\tilde{q}_i$, $i=1,2,3$ under the Cartan 
subgroup $[U(1)]^3$ of the $SO(6)$ R-symmetry   
takes the form  
\begin{equation}
\begin{split}
ds_{10}^2=&\sqrt{\triangle}\left[-(H_1H_2H_3)^{-1}fdt^2+\left(f^{-1}dr^2+r^2(d\calm_3)^2\right)\right]\\
&+\frac{1}{\sqrt{\triangle}}\sum_{i=1}^3L^2 H_i\left(d\mu_i^2+\mu_i^2\left[d\phi_i+a_i\ dt\right]^2\right)\,,
\end{split}
\eqlabel{metunc}
\end{equation}
where $\calm_3=\{R^3,S^3\}$ is a spatial manifold corresponding to curvatures $\k=\{0,1\}$, 
\begin{equation}
\begin{split}
&a_i=\frac{\tilde{q}_i}{q_i}\ L^{-1}\left(H_i^{-1}-1\right)\,,\qquad H_i=1+\frac{q_i}{r^2}\,, \\
&\triangle=H_1H_2H_3\sum_{i=1}^3\frac{\mu_i^2}{H_i}\,,\qquad f=\k-\frac{\mu}{r^2}+\frac{r^2}{L^2}H_1H_2H_3\,,
\end{split}
\end{equation}
and 
\begin{equation}
\mu_1=\cos\theta_1\,,\qquad \mu_2=\sin\theta_1\ \cos\theta_2\,,\qquad \mu_3=\sin\theta_1\ \sin\theta_2\,.
\eqlabel{defmu}
\end{equation}
The physical charges $\tq_i$ are related to charge parameters $q_i$ as  
\begin{equation}
\tilde{q}_i=\sqrt{q_i(\mu+\k q_i)}\,.
\end{equation}
Finally, the dilaton is constant, and the five form is given by 
\begin{equation}
F_5=\calf_5+\star \calf_5\,,\qquad \calf_5=d B_4\,,
\eqlabel{5form}
\end{equation}
with
\begin{equation}
B^{(4)}=-\frac{r^4}{L}\triangle\ dt\wedge dvol_{\calm_3}-L\sum_{i=1}^3 
\tilde{q}_i\mu_i^2\left(Ld\phi_i-\frac{q_i}{\tilde{q}_i}\ dt\right)
\wedge dvol_{\calm_3}\,,
\end{equation}
where $dvol_{\calm_3}$ is a volume form on $\calm_3$. 

Without loss of generality we can set $L=1$.
Computing $\a'$ corrections for the most general 3-charge black holes is rather complicated. 
In this paper we consider only the following special cases:
\nxt $\k=\{0,1\}$ and $q_i=0,\ i=1,2,3$;
\nxt $\k=\{0,1\}$ and $q_i=Q,\ i=1,2,3$.

Let $r=r_+$ be the position of the outer horizon of the black hole, \ie, the largest 
positive root of 
\begin{equation}
f(\rp)=0\,.
\eqlabel{horizon}
\end{equation} 
Neutral black holes  have the following thermodynamic properties
\begin{equation}
\begin{split}
&\pi T=\frac{\mu}{2\rp^3}+\frac{\rp}{2},\qquad S=\frac{n_c^2}{2\pi}\ \rp^3\,,\\
&E=\frac{n_c^2}{8\pi^2}\left(3\mu+\frac 34\k^2\right)\,,
\qquad F=\frac{n_c^2}{8\pi^2}\left(-\mu+\frac 34\k^2+2\rp^2\k\right)\,,
\end{split}
\eqlabel{neutral}
\end{equation}
where $T$ is the  Hawking temperature, $S$ is the entropy density, $E$ is the energy density,  and $F$ is the 
Helmholtz free energy density. Also,  the Hawking-Page phase transition occurs 
at temperature $T_{HP}$ when the black hole free energy
equals the ADM mass of the extremal $AdS_5\times S^5$ geometry \cite{w98}
\begin{equation}
F(\mu)=E(\mu=0) \qquad \Rightarrow\qquad \pi T_{HP}=\frac 32\ \k^{1/2}\,.
\eqlabel{thp}
\end{equation}  
 
For the charged black holes we have \cite{th1,th4}
\begin{equation}
\begin{split}
&\pi T=\frac{\mu}{2\rp^3}+\frac{\rp}{2}\ \frac{\rp^2-3Q}{\rp^2-Q}\,,\qquad S=\frac{n_c^2}{2\pi}\ \rp^3\,,\\
&E=\frac{n_c^2}{8\pi^2}\left(3\mu+\frac 34\kappa^2+6Q\kappa\right)\,,\qquad 
\Omega=\frac{n_c^2}{8\pi^2}\left(-\mu+\frac 34\kappa^2+2\rp^2\kappa-2Q\k\right)\,,\\
&\mu_q=\frac{n_c^2}{4\pi^2\rp^2}\ \left(\mu Q+\kappa Q^2\right)^{1/2}\,,\qquad \tQ=\sqrt{\k Q^2+\mu Q}\,,
\end{split}
\eqlabel{charged}
\end{equation}
where  $\Omega$ is the Gibbs free energy density, $\mu_q$ is the chemical potential corresponding to a physical 
charge density $\tQ$. Note that because all $U(1)$ R-symmetry charges are the same, the basic thermodynamic 
relation takes form   
\begin{equation}
\Omega=E-TS-3\mu_q \tQ\,,
\eqlabel{basic}
\end{equation}
while the first law of thermodynamics is
\begin{equation}
d\Omega=-S\ dT-3\tQ\ d\mu_q\,.
\eqlabel{1stlaw}
\end{equation}

\section{Computational framework}
We start with the  tree level type IIB low-energy effective action in ten dimensions taking into account the 
leading order string corrections\footnote{In \eqref{aaa} ellipses stand for other fields not essential for the present analysis.} 
\cite{cor1,cor2,cor3,cor4}
\begin{equation}
I=  \frac{1}{ 16\pi G_{10}} \int_{\calm_{10}} d^{10} x \sqrt {-g}
\ \bigg[ R_{10} - {\frac 12} (\partial \phi)^2 - \frac{1}{4 \cdot 5!}   (F_5)^2  +...+ 
\  \gamma \ e^{- {\frac 3 2} \phi}  W + ...\bigg]   \  ,
\eqlabel{aaa}
\end{equation}
\begin{equation}
 \ \ \ \ \ \   
  \gamma= { \frac 18} \zeta(3)(\alpha')^3 \ , 
\eqlabel{defg}
\end{equation}
where 
\begin{equation}
W =  C^{hmnk} C_{pmnq} C_{h}^{\ rsp} C^{q}_{\ rsk} 
 + {\frac 12}  C^{hkmn} C_{pqmn} C_h^{\ rsp} C^{q}_{\ rsk}\  . 
\eqlabel{rrrr}
\end{equation}
As in \cite{gkt,bls} we assume that in a chosen scheme self-dual $F_5$ form does not receive 
order $(\a')^3$
corrections. In \cite{bb} it was shown that such assumption leads to consistent description 
of hydrodynamic fluctuations of the $\k=0$ nonextremal black holes, which following gauge theory-string theory 
correspondence describes transport properties of strongly coupled finite temperature $\caln=4$ Yang-Mills plasma 
at finite 't Hooft coupling.  
Specifically, the shear diffusion pole of the two-point correlation function of plasma stress-energy tensor is insensitive 
to the matter part (flux terms) in effective action \eqref{aaa} (for a general discussion see \cite{univ}). On the other hand, 
the sound pole in the stress-energy two-point correlation function depends on the matter content of the gravitational 
action (compare \cite{bbs} and \cite{bhyd}), and thus is expected to be sensitive to the $(\a')^3$ structure of the 
flux terms in \eqref{aaa}. Now, the shear viscosity computed from the diffusive pole must agree with the shear viscosity extracted 
from the attenuation of the sound waves. This agreement of both quantities derived from a specific form of the $\a'$-corrected effective 
action \eqref{aaa} has been emphasized in \cite{bb}\footnote{It would be interesting to test consistency 
of black hole thermodynamics and hydrodynamics in $\a'$ -corrected type IIB supergravity effective action advocated in
\cite{rc6,rc7,rrcc}.}.

We represent  ten dimensional background geometry describing $\gamma$-corrected charged black  
holes by the following ansatz
\begin{equation}
\begin{split}
ds_{10}^2=&-D_1^2\ dt^2+D_2^2\   \frac {1}{\k}(dS^3)^2+D_3^2\ dr^2+D_4^2\ d\theta_1^2+D_4^2\ \sin^2\theta_1 d\theta_2^2\\
&+D_4^2\ \sum_{i=1}^3\ \mu_i^2\biggl(d\phi_i+Da\ dt\biggr)^2\,,
\end{split}
\eqlabel{g10b}
\end{equation}
where $D_i=D_i(r,\theta_1,\theta_2)$ and $Da=Da(r, \theta_1,\theta_2)$, $\mu_i$  are given by \eqref{defmu},  and $\left(dS^3\right)^2$ is a metric 
on a round three-sphere of unit radius. Taking $\k\to 0$ limit in \eqref{g10b} we obtain charged (or in ten dimensions 
rotating ) black hole solution 
with $\k=0$ spatial curvature of the horizon. Notice that in writing the metric ansatz \eqref{g10b} 
we assumed that $[U(1)]^3$ isometry is not broken by $\a'$ corrections, also we assumed that our black hole carries 
identical charges under these $U(1)$ symmetries. 
We further assume that the dilaton $\phi=\phi(r,\theta_1,\theta_2)$. Since in the supergravity approximation 
$\phi\biggl|_{\a'=0}=0$ we see that $\phi\propto \g$, thus, as in \cite{gkt},  to leading order in $\g$ 
the metric deformations and the dilaton would decouple.  

Next, we would like to obtain effective action involving supergravity modes $D_i$, $Da$, and $\phi$.
To achieve this\footnote{Effective low-dimensional supergravity actions derived in this 
way were recently used in \cite{b1,aby}.}, one needs to evaluate the $\a'$ string corrected action \eqref{aaa} on the metric 
ansatz \eqref{g10b}. The latter is a straightforward though a tedious exercise for a ten dimensional Ricci 
scalar and the fourth order Weyl tensor invariant \eqref{rrrr}\footnote{Because of their complexity, 
we do not present  explicit expressions for \eqref{rmetric}.} 
\begin{equation}
\begin{split}
&R_{10}= R_{10}\biggl[D_i(r,\theta_1,\theta_2),\ Da(r,\theta_1,\theta_2),\ \{\theta_1,\theta_2\}\biggr]\,,\\
&W= W\biggl[D_i(r,\theta_1,\theta_2),\ Da(r,\theta_1,\theta_2),\ \{\theta_1,\theta_2\}\biggr]\,.
\end{split}
\eqlabel{rmetric}
\end{equation}
A special care has to be taken 
with the five-form contribution  in \eqref{aaa} \cite{gd}. In the latter case we find that integrating out the 5-form 
one gets a contribution
\begin{equation}
- \frac{1}{4 \cdot 5!}   (F_5)^2=-\frac{4Q(\mu+\kappa Q)}{D_4^4D_2^6}-\frac{8}{D_4^{10}}\,.
\eqlabel{5formcont}
\end{equation}
We separate 3-dimensional effective action describing $\g$-corrected metric 
derived from \eqref{aaa} into contribution coming from  purely supergravity part
and the $\g$-correction  
\begin{equation}
S_{eff}=S_{SUGRA}+S_{\g}\,,
\eqlabel{seffsugra}
\end{equation}
with 
\begin{equation}
\begin{split}
S_{SUGRA}=&\frac{vol(S^3)}{\k^{3/2}}\ \frac{(2\pi)^3}{16\pi G_{10}}\int dt\ \times \int_0^{\pi/2}d\theta_1\int_{0}^{\pi/2}d\theta_2
 \int^{\infty}_{\rp}dr\ \sin^3\theta_1\cos\theta_1\sin\theta_2\cos\theta_2\\
&\times \ D_1D_2^3D_3D_4^5\ \biggl\{R_{10}-\frac{4Q(\mu+\kappa Q)}{D_4^4D_2^6}-\frac{8}{D_4^{10}}\biggr\}\,,
\end{split}
\eqlabel{sugrapart}
\end{equation}  
and
\begin{equation}
\begin{split}
S_{\g}=&\frac{vol(S^3)}{\k^{3/2}}\ \frac{(2\pi)^3}{16\pi G_{10}}\int dt\ \times \int_0^{\pi/2}d\theta_1\int_{0}^{\pi/2}d\theta_2
\int^{\infty}_{\rp}dr\ \sin^3\theta_1\cos\theta_1\sin\theta_2\cos\theta_2\\
&\times \ D_1D_2^3D_3D_4^5\ \biggl\{\g\ W\biggr\}\,.
\end{split}
\eqlabel{gpart}
\end{equation}  
In \eqref{sugrapart} and \eqref{gpart} 
$\rp$ is the position of the outer horizon of metric \eqref{g10b} determined as the largest positive root of 
\begin{equation}
D_1(\rp,\theta_1,\theta_2)=0\,.
\end{equation}
We find that even including $\a'$ corrections $\rp$ is actually independent of $\theta_i$. 
Lastly, the (decoupled to leading order in $\g$ ) equation for the  dilaton reads
\begin{equation}
\square \phi=\g\ \frac 32 W\,, 
\eqlabel{dilaton}
\end{equation}
where both the Dalambertian and the fourth order Weyl tensor invariant can be evaluated in 
$\a'$ uncorrected metric \eqref{metunc}.

Effective action \eqref{seffsugra} provides a consistent ``Kaluza-Klein'' reduction of \eqref{aaa} on 
$R\times \calm_3\times [S^1]^3$ (along the time direction, world-volume $R^3$ or $S^3$ directions, 
and $[U(1)]^3$ isometries parameterized by $\phi_i$) in a sense that any solution of \eqref{seffsugra}
is a solution of \eqref{aaa} \cite{b1}. Equations of motion derived from 
\eqref{seffsugra} take the form
\begin{equation}
\begin{split}
&0=\frac{\dd S}{\dd D_i}+\g J_i\,,\qquad 0=\frac{\dd S}{\dd Da}+\g J_a\,,\\
&\g J_i\equiv \frac{\dd S_{\g}}{\dd D_i}\,,\qquad \g J_a\equiv \frac{\dd S_{\g}}{\dd Da}\,.
\end{split}
\eqlabel{eom}
\end{equation}
Explicit evaluation of \eqref{eom} is the most computationally consuming part of our framework. 
Again, due to their complexity, explicit expressions for  \eqref{eom} will not be presented here.

To order $\g^0$ \eqref{eom} can be solved with (in agreement with \eqref{metunc})
\begin{equation}
\begin{split}
&D_1^{(0)}=\frac{r^2}{r^2+Q}\ \left(\k-\frac{\mu}{r^2}+r^2\left(1+\frac{Q}{r^2}\right)^3\right)^{1/2}\,,\qquad 
D_2^{(0)}=(r^2+Q)^{1/2}\,,\\
&D_3^{(0)}=\left(1+\frac{Q}{r^2}\right)^{1/2}\ \left(\k-\frac{\mu}{r^2}+r^2\left(1+\frac{Q}{r^2}\right)^3\right)^{-1/2}\,,
\qquad D_4^{(0)}=1\,,\\
&Da^{(0)}=-\frac{(Q\mu+\k^2 Q)^{1/2}}{r^2+Q}\,.
\end{split}
\eqlabel{zeroorder}
\end{equation}
To order $\g^1$ we find it convenient to choose a  radial coordinate so that 
\begin{equation}
\begin{split}
&D_1(r,\theta_1,\theta_2)=D_1^{(0)}\times \left(1+\g\ \left(A+B-\frac 53 \nu\right)\right)\,,\\
&D_2(r,\theta_1,\theta_2)=D_2^{(0)}\times \left(1+\g\ \left(A-\frac 53 \nu\right)\right)\,,\\
&D_3(r,\theta_1,\theta_2)=D_3^{(0)}\times \left(1+\g\ \left(2A-B-\frac 53 \nu\right)\right)\,,\\
&D_4(r,\theta_1,\theta_2)=D_4^{(0)}\times \left(1+\g\ \nu\right)\,,\qquad Da(r,\theta_1,\theta_2)=Da^{(0)} 
\times \left(1+\g\ a\right) \,.
\end{split}
\eqlabel{1order}
\end{equation}
Upon deriving equations of motion \eqref{eom} in their generality, we find that it is consistent to assume 
that deformations $A, B, \nu, a$  are functions of the radial coordinate only. In other words, given \eqref{1order}
and \eqref{eom} we obtain second order  system of ordinary differential equations 
for $\{A,B,\nu,a\}$.  The advantage of fixing radial coordinate 
to order $\g$ as in \eqref{1order} is that differential equations for each deformation factor decouple with such a choice.  
 
We conclude this section by specifying the boundary conditions for $\{A,B,\nu,a\}$: 
\begin{equation}
\begin{split} 
&\{A,B,\nu,a\}\qquad {\rm are\ regular\ functions\ as\ } \qquad r\to\rp+0 \,,\\
&\{A,B,\nu,a\}\to 0 \qquad {\rm  as\ } \qquad r\to\infty\,.
\end{split}
\eqlabel{bc}
\end{equation}

\section{$\a'$-correct black hole solutions in asymptotic $AdS_5\times S^5$ type IIB supergravity}

In this section we present differential equations (and their) solutions describing 
$\a'$ corrections to neutral and charged black holes with $\k=\{0,1\}$ spatial 
curvature of the horizon in asymptotic $AdS_5\times S^5$ type IIB supergravity. 
The equations are obtained from \eqref{eom} within ansatz
\eqref{1order}. We find that solutions to these equations  subject to boundary conditions \eqref{bc} 
are unique. It is simplest to write these equations using a new radial coordinate
\begin{equation}
x\equiv r^2+Q\,.
\eqlabel{xdef}
\end{equation}
We also denote 
\begin{equation}
\xp\equiv \rp^2+Q\,.
\end{equation}

\subsection{Neutral black hole with horizon spatial curvature $\k$}\label{neutralf}
In this case the decoupled system of equations is the simplest. 
We find\footnote{The prime denotes derivative with respect to $x$.}
\begin{equation}
0=A''+ \frac{360 \mu^3}{x^8}\,,
\eqlabel{Ameq}
\end{equation}
\begin{equation}
\begin{split}
0=&B''+ \frac{9 x^2-\mu+5 \k x}{2x (x^2+\k x-\mu)}\ B'+ \frac {12 x^2-2 \mu+7 \k x}{2x (x^2+\k x-\mu)}\ A'
+ \frac{6 x+\k}{2x (x^2+\k x-\mu)}\ B\\
&-\frac{12 x+\k}{2x (x^2+\k x-\mu)}\ A -\frac{5 \mu^3 (-639 \mu+608 \k x+576 x^2)}{
2x^8 (x^2+\k x-\mu)}\,,
\end{split}
\eqlabel{Bmeq}
\end{equation}
\begin{equation}
\begin{split}
0=\nu''+\frac{2 \k x-\mu+3 x^2}{x (x^2+\k x-\mu)}\ \nu'-\frac{8}{x^2+\k x-\mu}\ \nu+\frac{135 \mu^4}{8x^8 (x^2+\k x-\mu)}\,.
\end{split}
\eqlabel{numeq}
\end{equation}
Additionally, there is a first order constrain arising from reparametrization invariance of the radial coordinate
\begin{equation}
\begin{split}
0=B'+\frac{4 x^2-2 \mu+3 \k x}{x^2+\k x-\mu}\ A'+\frac{\k+2 x}{x^2+\k x-\mu}\ B-\frac{\k+4 x}{x^2+\k x-\mu}\ A
+ \frac{5 \mu^3 (16 \k x-27 \mu)}{x^7 (x^2+\k x-\mu)}\,.
\end{split}
\eqlabel{cmeq}
\end{equation}
It is straightforward to verify that \eqref{cmeq} is consistent with \eqref{Ameq} and \eqref{Bmeq}.
We can explicitly solve for $A$ and $B$ subject to boundary conditions \eqref{bc}
\begin{equation}
A=-\frac{60\mu^3}{7x^6}\,,
\eqlabel{asol}
\end{equation}
\begin{equation}
B=\frac{\mu^3}{x^2+\k x-\mu}\ \left(\frac{60}{x^4}+\frac{340\k}{7x^5}-\frac{555\mu}{14x^6}+\calb\right)\,,
\eqlabel{bsol}
\end{equation}
where 
\begin{equation}
\calb=-\frac{60}{\xp^4}-\frac{340\k}{7\xp^5}+\frac{555\mu}{14\xp^6}\,.
\eqlabel{bb}
\end{equation}

Unfortunately, we were unable to solve \eqref{numeq} apart from $\k=0$ case. In the latter case we reproduce the result of 
\cite{pt}
\begin{equation}
\nu\bigg|_{\k=0}\equiv \nu_0=\frac{15\mu^2(\mu+x^2)}{32x^6}\,.
\eqlabel{nu0}
\end{equation}
Notice that in the limit $\k\to 0$ and with the radial coordinate reparametrization 
\begin{equation}
r\to r\left(1+\g\ \frac {60\rp^{12}}{7r^{12}}\right)\qquad {\rm and}\qquad  \rp\to \rp\left(1+\g\ \frac {60}{7}\right)\,,
\eqlabel{repam}
\end{equation}
our solution reproduces the known $\a'$-corrected geometry of the  near-extremal flat D3-branes \cite{gkt,pt}.
In is straightforward to analyze equation \eqref{numeq} and observe that the absence of the singularity 
at $x=\xp$ along with the ultraviolet (UV) boundary conditions $\nu\to 0$ as $x\to \infty$, uniquely 
determine the $S^5$ warp factor $\nu(x)$. Turns out that for the thermodynamics of the neutral black holes 
we  would only need the UV asymptotic of $\nu$ as $x\to \infty$
\begin{equation}
\nu(x)\sim \calo(x^{-4})\,.
\eqlabel{nuass}
\end{equation}

For completeness we write down the solution to the dilaton equation \eqref{dilaton}
\begin{equation}
\phi=\frac{45\mu^4}{4}\int_x^{\infty}\ \frac{dz}{z(z^2+\k z-\mu)}\ \left(\frac{1}{z^6}-\frac{1}{\xp^6}\right)\,.
\eqlabel{dils}
\end{equation}
Integral in \eqref{dils} can be evaluated explicitly. It is easy to see that the $\k=0$ limit reproduces 
the result of \cite{gkt,pt}.

We conclude this section with comments on the $\a'$ corrections to global extremal $AdS_5\times S^5$ geometry. 
From previous work \cite{gkt,pt} we know that the fourth order Weyl invariant \eqref{rrrr} vanishing at the extremality in the 
Poincare patch. Since it is a local invariant, it must vanish in global coordinates as well. The latter implies 
that the global extremal $AdS_5\times S^5$ geometry does not receive $\a'$ corrections\footnote{Actually, Kallosh-Rajaraman
arguments \cite{kr} are much more powerful than $\a'$-leading order explicit computations and imply that 
global extremal $AdS_5\times S^5$ geometry does not receive $\a'$ corrections at any order.}. From \eqref{asol}, \eqref{bsol}
we see that at the extremality, \ie, setting $\mu=0$ 
\begin{equation}
A\bigg|_{\mu=0}=0\,,\qquad B\bigg|_{\mu=0}=0\,.
\eqlabel{extrem}
\end{equation}
Furthermore, the general solution of \eqref{numeq} at the extremality reads
\begin{equation}
\begin{split}
\nu\bigg|_{\mu=0}=&\calc_1 (3 \k^2+12 \k x+10 x^2)\\
&+\frac{\calc_2}{x}\ 
 \left(-3 x (3 \k^2+12 \k x+10 x^2) \ln\left(1+\frac \k x\right)+\k (\k^2+21 \k x+30 x^2\right)\,.
\end{split}
\eqlabel{gennu}
\end{equation}
Now, the appropriate boundary conditions is that $\nu$ is nonsingular as $x\to +0$, and vanishes as $x\to \infty$.
Thus
\begin{equation}
\nu\bigg|_{\mu=0}=0\,.
\end{equation}

\subsection{$\k=1$ charged black holes}
In this case we find
\begin{equation}
\begin{split}
0=&A''+J_A\,,
\end{split}
\eqlabel{Ameq1}
\end{equation}
\begin{equation}
\begin{split}
0=&B''-\frac{2 (\mu-2 x-3 x^2+2 Q)}{x^2-2 x Q+Q^2+x^3-\mu x+Q \mu}\ B'
+ \frac{2 (1+3 x)}{x^2-2 x Q+Q^2+x^3-\mu x+Q \mu}\ B\\
&-\frac{4 (\mu-2 x-3 x^2+2 Q)}{x^2-2 x Q+Q^2+x^3-\mu x+Q \mu}\ A'-\frac{2 (1+6 x)}{x^2-2 x Q+Q^2+x^3-\mu x+Q \mu}\ A+J_B\,,
\end{split}
\eqlabel{Bmeq1}
\end{equation}
\begin{equation}
\begin{split}
&0=\nu''-\frac{\mu-2 x-3 x^2+2 Q}{x^2-2 x Q+Q^2+x^3-\mu x+Q \mu} \nu' 
-\frac{4 (Q^2+Q \mu+10 x^3)}{5x^2 (x^2-2 x Q+Q^2+x^3-\mu x+Q \mu)}\nu\\
& -\frac{3 (Q^2+Q \mu+4 x^3-2 \mu x-4 x Q+3 x^2)}{
5x (x^2-2 x Q+Q^2+x^3-\mu x+Q \mu)}\ A'-\frac{3}{5x}\ B'\\
&+\frac{3 (Q^2+Q \mu+x^2+4 x^3)}{5x^2 (x^2-2 x Q+Q^2+x^3-\mu x+Q \mu)}\ A+ \frac{3 (Q^2+Q \mu-2 x^3-x^2)}{5
x^2 (x^2-2 x Q+Q^2+x^3-\mu x+Q \mu)}\ B\\
&+J_{\nu}\,,
\end{split}
\eqlabel{numeq1}
\end{equation}
\begin{equation}
\begin{split}
0=&a''-\frac{16}{3x}\ \nu'+J_a\,.
\end{split}
\eqlabel{aeq1}
\end{equation}
Additionally, there is a first order constraint arising from reparametrization invariance of the radial coordinate
\begin{equation}
\begin{split}
&0=(Q^2+Q \mu+4 x^3-2 \mu x-4 x Q+3 x^2)\ A'+(x^2-2 x Q+Q^2+x^3-\mu x+Q \mu)\ B'
 \\
&- \frac{Q (Q+\mu)}{3}\ a'-\frac{Q^2+Q \mu+x^2+4 x^3}{x }
 \ A -\frac{Q^2+Q \mu-2 x^3-x^2}{x }\ B+\frac{16 Q (Q+\mu)}{9 x }\ \nu \\
&+ \frac{Q (Q+\mu)}{3x }\ a+J_{const}\,.
\end{split}
\eqlabel{cmeq1}
\end{equation}
Explicit expressions for the inhomogeneous sources $\{J_A, J_B, J_{\nu}, J_a, J_{const}\}$ are given in Appendix \ref{k1}.
It is straightforward to verify that \eqref{cmeq1} is consistent with \eqref{Ameq1}-\eqref{aeq1}.
Furthermore, equations \eqref{Ameq1}-\eqref{cmeq1} in the limit $Q\to 0$ reproduce equations 
\eqref{Ameq}-\eqref{cmeq} in the limit $\k\to 1$.

Solution of \eqref{Ameq1} with boundary conditions \eqref{bc} is 
\begin{equation}
\begin{split}
A=&\frac{27569Q^3}{360x^9} (\mu^3+Q^3+3 \mu Q^2+3 \mu^2 Q)-\frac{21049Q^2}{216x^8}  
(\mu^3+2 Q^3+5 \mu Q^2+4 \mu^2 Q)\\
&+\frac{Q}{112x^7}  (5359 \mu^3+23799 \mu^2 Q+36880 \mu Q^2+18440 Q^3)+\frac{1}{x^6}\biggl(
-\frac{320}{7} Q^3+\frac{38}{21} \mu^2 Q^2\\
&+\frac{38}{21} Q^4-\frac{480}{7} \mu Q^2-\frac{60}{7} 
\mu^3-40 \mu^2 Q+\frac{76}{21} Q^3 \mu\biggr)-\frac{4Q}{9x^5}  (2 Q^2+3 Q \mu+\mu^2)\,.
\end{split}
\eqlabel{as1}
\end{equation}
The most general solution of \eqref{Bmeq1} is 
\begin{equation}
\begin{split}
&B=\frac{x \calb_1+\calb_2}{x^3+x^2-(\mu+2 Q) x+Q (Q+\mu)}+\frac{1}{15120 x^9(x^3+x^2-(\mu+2 Q) x+Q (Q+\mu))}\\
&\times \biggl(
-302400 Q (Q+\mu) (2 Q+\mu) x^7+( 554400 Q^4+4005120 Q^3+1108800 Q^3 \mu\\
&+554400 \mu^2 Q^2+6007680 \mu Q^2+3816960 \mu^2 Q+907200 \mu^3) x^6+( -14110176 Q^4\\
&-28220352 Q^3 \mu+3594240 Q^3
+5391360 \mu Q^2-18524976 \mu^2 Q^2+3265920 \mu^2 Q\\
&-4414800 \mu^3 Q+734400 \mu^3) x^5+( 16345800 Q^5+40864500 Q^4 \mu-18626040 Q^4\\
&-37252080 Q^3 \mu+32691600 Q^3 \mu^2-26330625 \mu^2 Q^2+8172900 \mu^3 Q^2-7704585 \mu^3 Q\\
&-599400 \mu^4) x^4 -10 Q (Q+\mu) (629580 Q^4+1259160 Q^3 \mu-3870674 Q^3-5806011 \mu Q^2\\
&+629580 \mu^2 Q^2-2679043 \mu^2 Q-371853 \mu^3) x^3 -2 Q^2 (20176669 Q^2+20176669 Q \mu\\
&+4556605 \mu^2) (Q+\mu)^2 x^2+ 10549266 Q^3 (2 Q+\mu) (Q+\mu)^3 x-4423125 Q^4 (Q+\mu)^4
\biggr)\,,
\end{split}
\eqlabel{bs1}
\end{equation}
where $\calb_i$ are integration constants. Given \eqref{as1} and \eqref{bs1} we can determine 
$a(x)$ from \eqref{cmeq1}
\begin{equation}
\begin{split}
&a=\frac{3 \calb_2}{Q (Q+\mu)}+x \left(\calc_a+\frac{16}{3}\  \int_{\xp}^{x}dz\ \frac{\nu(z)}{z^2}\right)
-\frac{1}{2520x^9} \biggl(-60480 Q (Q+\mu) x^6\\
&+(628992 Q^2+241920 \mu^2+628992 Q \mu) x^5+( 564480 Q \mu-1404480 Q^3-2106720 \mu Q^2\\
&-702240 \mu^2 Q+564480 Q^2+201600 \mu^2) x^4+( -2093760 Q^3+763920 Q^4-151200 \mu^3\\
&+763920 \mu^2 Q^2-3140640 \mu Q^2-1349280 \mu^2 Q+1527840 Q^3 \mu) x^3\\
&+315 Q (Q+\mu) (1909 \mu^2+9372 Q \mu+9372 Q^2) x^2 -931000 Q^2 (2 Q+\mu) (Q+\mu)^2 x\\
&+440622 Q^3 (Q+\mu)^3\biggr)\,,
\end{split}
\eqlabel{ags1}
\end{equation}
where $\calc_a$ is a new integration constant. 
Since the boundary conditions are such that $a\to 0$ as $x\to \infty$, we conclude from \eqref{ags1} 
that 
\begin{equation}
\calb_2=0\,,\qquad \calc_a=-\frac{16}{3}\ \int_{\xp}^{\infty}dz\ \frac{\nu(z)}{z^2}\,.
\eqlabel{b2ca}
\end{equation}
The remaining integration constant $\calb_1$ is fixed requiring that $B$ is nonsingular as $x\to\xp+0$
\begin{equation}
\begin{split}
&\calb_1=\frac{1}{15120\xp^{10}}\biggl(302400 Q (2 Q+\mu) (Q+\mu) \xp^7+(-554400 Q^4-4005120 Q^3-907200 \mu^3\\
&-1108800 Q^3 \mu-6007680 \mu Q^2-3816960 \mu^2 Q-554400 \mu^2 Q^2) \xp^6+( -3594240 Q^3\\
&-734400 \mu^3+14110176 Q^4+28220352 Q^3 \mu+18524976 \mu^2 Q^2-3265920 \mu^2 Q\\
&+4414800 \mu^3 Q-5391360 \mu Q^2) \xp^5+( -16345800 Q^5+599400 \mu^4+18626040 Q^4\\
&-40864500 \mu Q^4+7704585 \mu^3 Q-8172900 \mu^3 Q^2+37252080 Q^3 \mu+26330625 \mu^2 Q^2\\
&-32691600 \mu^2 Q^3) \xp^4 -10 Q (Q+\mu) (371853 \mu^3+2679043 \mu^2 Q-629580 \mu^2 Q^2\\
&-1259160 Q^3 \mu+5806011 \mu Q^2-629580 Q^4+3870674 Q^3) \xp^3+ 2 Q^2 (4556605 \mu^2\\
&+20176669 Q \mu+20176669 Q^2) (Q+\mu)^2 \xp^2-10549266 Q^3 (2 Q+\mu) (Q+\mu)^3 \xp\\
&+4423125 Q^4 (Q+\mu)^4\biggr)\,.
\end{split}
\eqlabel{b1}
\end{equation}

The resulting equation  for the $S^5$ warp factor $\nu(x)$ is
\begin{equation}
\begin{split}
&0=\nu''-\frac{\mu-2 x-3 x^2+2 Q}{x^2-2 x Q+Q^2+x^3-\mu x+Q \mu} \nu'-\frac{4(Q^2+Q \mu+10 x^3)}{5x^2 (x^2-2 x Q+Q^2+x^3-\mu x+Q \mu)} \nu\\
&+\cali_\nu\,,
\end{split}
\eqlabel{final1}
\end{equation}
where $\cali_{\nu}$ is given in Appendix \ref{k1}.
We were unable to find analytic solution to \eqref{final1} even for\footnote{In the case 
of $\k=0$ charged black holes the corresponding equation can be solved perturbatively in $Q$.} 
$Q=0$. Nonetheless, it is straightforward to verify 
that requiring nonsingularity at the horizon and the vanishing of $\nu(x)$ as $x\to \infty$, the $S^5$ warp factor $\nu(x)$
is uniquely determined. As in  \eqref{nuass}
\begin{equation}
\nu(x)\sim \calo(x^{-4})\,.
\eqlabel{nuass1}
\end{equation}

For completeness we present the dilaton equation 
\begin{equation}
\begin{split}
0=\phi''-\frac{-3 x^2+\mu+2 Q-2 x}{-\mu x+x^3+Q^2-2 Q x+x^2+Q \mu}\ \phi'+J_\phi\,,
\end{split}
\eqlabel{dil1}
\end{equation}
where $J_\phi$ is given in Appendix \ref{k1}.
The dilaton equation can be solved analytically.

\subsection{$\k=0$ charged black branes }\label{chargedf} 
In this case we find
\begin{equation}
\begin{split}
0=&A''+J_A\,,
\end{split}
\eqlabel{Ameq0}
\end{equation}
\begin{equation}
\begin{split}
0=&B''+\frac{2 (3 x^2-\mu)}{-x \mu+Q \mu+x^3}\ B'+ \frac{6 x}{-x \mu+Q \mu+x^3}\ B
+\frac{4 (3 x^2-\mu)}{-x \mu+Q \mu+x^3}\ A'\\
&-\frac{12 x}{-x \mu+Q \mu+x^3}\ A+J_B\,,
\end{split}
\eqlabel{Bmeq0}
\end{equation}
\begin{equation}
\begin{split}
0=&\nu''+\frac{3 x^2-\mu}{-x \mu+Q \mu+x^3}\ \nu'-\frac{4 (Q \mu+10 x^3)}{5 x^2 (-x \mu+Q \mu+x^3)}\ \nu
 -\frac{3(4 x^3+Q \mu-2 x \mu)}{5x (-x \mu+Q \mu+x^3)}\ A'\\
&-\frac{3}{5x}\ B'+ \frac{3 (Q \mu+4 x^3)}{5x^2 (-x \mu+Q \mu+x^3)}
\ A+ \frac{3 (Q \mu-2 x^3)}{5x^2 (-x \mu+Q \mu+x^3)}\ B+J_{\nu}\,,
\end{split}
\eqlabel{numeq0}
\end{equation}
\begin{equation}
\begin{split}
0=&a''-\frac{16}{3x}\ \nu'+J_a\,.
\end{split}
\eqlabel{aeq0}
\end{equation}
Additionally, there is a first order constraint arising from reparametrization invariance of the radial coordinate
\begin{equation}
\begin{split}
0=&(4 x^3+Q \mu-2 x \mu)\ A'+(-x \mu+Q \mu+x^3)\ B'-\frac{ Q \mu}{3}\ a'-\frac{Q \mu+4 x^3}{x}\ A
\\
& -\frac{Q \mu-2 x^3}{x}\ B + \frac{16 \mu Q}{9x}\ \nu+  \frac{\mu Q}{3x}\ a+J_{const}\,.
\end{split}
\eqlabel{cmeq0}
\end{equation}
Explicit expressions for the inhomogeneous sources $\{J_A, J_B, J_{\nu}, J_a, J_{const}\}$ are given in Appendix \ref{k0}.
It is straightforward to verify that \eqref{cmeq0} is consistent with \eqref{Ameq0}-\eqref{aeq0}.
Furthermore, equations \eqref{Ameq0}-\eqref{cmeq0} in the limit $Q\to 0$ reproduce equations 
\eqref{Ameq}-\eqref{cmeq} in the limit $\k\to 0$.

Solution of \eqref{Ameq0} with boundary conditions \eqref{bc} is 
\begin{equation}
\begin{split}
A=&\frac{27569\mu^3 Q^3}{360x^9} -\frac{21049Q^2 \mu^3}{216x^8}+\frac{5359Q \mu^3}{112x^7}
+\frac{2\mu^2 (19 Q^2-90 \mu)}{21x^6}
-\frac{4 Q \mu^2}{9x^5}\,.
\end{split}
\eqlabel{as0}
\end{equation}
The most general solution of \eqref{Bmeq0} is 
\begin{equation}
\begin{split}
&B=\frac{x \calb_1+\calb_2}{x^3-\mu x+Q \mu}+\frac{1}{15120 x^9 (x^3-\mu x+Q \mu)} \biggl(
-4423125 \mu^4 Q^4+10549266 Q^3 \mu^4 x\\
&-9113210 Q^2 \mu^4 x^2-30 (-123951 \mu+209860 Q^2) \mu^3 Q x^3+8100 (1009 Q^2-74 \mu) \mu^3 x^4\\
&-4414800 Q \mu^3 x^5+50400 \mu^2 (11 Q^2+18 \mu) x^6-302400 Q \mu^2 x^7\biggr)\,,
\end{split}
\eqlabel{bs0}
\end{equation}
where $\calb_i$ are integration constants. Given \eqref{as0} and \eqref{bs0} we can determine 
$a(x)$ from \eqref{cmeq0}
\begin{equation}
\begin{split}
&a=\frac{3 \calb_2}{Q \mu}+x \left(\calc_a+\frac{16}{3} \int_{\xp}^{x}dz\ \frac{\nu(z)}{z^2}\right)
-\frac{\mu}{2520x^9} \biggl( 440622 Q^3 \mu^2-931000 Q^2 \mu^2 x\\
&+601335 Q \mu^2 x^2+(763920 \mu Q^2
-151200 \mu^2) x^3-702240 Q \mu x^4+241920 \mu x^5\\
&-60480 Q x^6\biggr)\,,
\end{split}
\eqlabel{ags0}
\end{equation}
where $\calc_a$ is a new integration constant. 
Since the boundary conditions are such that $a\to 0$ as $x\to \infty$, we conclude from \eqref{ags0} 
that 
\begin{equation}
\calb_2=0\,,\qquad \calc_a=-\frac{16}{3}\ \int_{\xp}^{\infty}dz\ \frac{\nu(z)}{z^2}\,.
\eqlabel{b2ca0}
\end{equation}
The remaining integration constant $\calb_1$ is fixed requiring that $B$ is nonsingular as $x\to\xp+0$
\begin{equation}
\begin{split}
&\calb_1=\frac{\mu^2}{15120\xp^{10}} \biggl( 302400 Q \xp^7+(-554400 Q^2-907200 \mu) \xp^6+ 4414800 Q \mu \xp^5\\
&-8100 \mu (1009 Q^2-74 \mu) \xp^4+30 Q \mu (-123951 \mu+209860 Q^2) \xp^3+ 9113210 Q^2 \mu^2 \xp^2\\
&-10549266 Q^3 \mu^2 \xp+4423125 Q^4 \mu^2\biggr)\,.
\end{split}
\eqlabel{b10}
\end{equation}

The resulting equation  for the $S^5$ warp factor $\nu(x)$ is
\begin{equation}
\begin{split}
&0=\nu''+\frac{3 x^2-\mu}{-x \mu+Q \mu+x^3}\ \nu'-\frac{4 (Q \mu+10 x^3)}{5x^2 (-x \mu+Q \mu+x^3)}\ \nu
+\cali_{\nu}\,,
\end{split}
\eqlabel{final0}
\end{equation}
where $\cali_{\nu}$ is given in Appendix \ref{k0}.
We were unable to find analytic solution to \eqref{final0} for general $Q$. In the next section we solve this equation analytically to 
leading order in $Q$. 
It is straightforward to verify 
that requiring nonsingularity at the horizon and the vanishing of $\nu(x)$ as $x\to \infty$, the $S^5$ warp factor $\nu(x)$
is uniquely determined. Again, as in  \eqref{nuass}
\begin{equation}
\nu(x)\sim \calo(x^{-4})\,.
\eqlabel{nuass0}
\end{equation}

For completeness we present the dilaton equation 
\begin{equation}
\begin{split}
0=\phi''-\frac{-3 x^2+\mu}{Q \mu+x^3-\mu x}\ \phi'+J_{\phi}\,,
\end{split}
\eqlabel{dil0}
\end{equation}
where $J_\phi$ is given in Appendix \ref{k0}.
The dilaton equation can be solved analytically.

\subsubsection{$\k=0$ charge black branes at large temperature}
As we explain in Section \ref{thermo}, to analyze the thermodynamics of  charged black holes
one needs to know explicitly the value  of the gauge potential $a(x)$ at the horizon, $x=\xp$. 
From \eqref{ags0}, the latter implies that we need to know explicit expression for  $\calc_a$ 
\eqref{b2ca0}. We were unable to compute $\calc_a$ for arbitrary values of the nonextremality 
parameter $\mu$ and the charge parameter $Q$. In this section we evaluate $\calc_a$ 
to leading order in $Q$. Physically, above approximation means that we would be interested in 
the thermodynamics of $\k=0$ charged black holes in the regime when the physical 
charge chemical potential 
$\mu_q\ll T$. 
 
To leading order in $Q$ the solution to \eqref{final0} which is nonsingular at the horizon $x=\xp$ and 
vanishes as $x\to \infty$ is given by
\begin{equation}
\begin{split}
\nu=&\nu_0+Q\biggl(-\frac{113241(2 x^2-\mu)}{5600\mu^{3/2}} \left(
\ln\left(1-\frac{\mu}{x^2}\right)+2 {\rm arctanh} \frac {\sqrt{\mu}}{x}\right) + \frac{9857\mu^3}{3920x^7}
-\frac{1144277 \mu^2}{98000x^5}\\
&+\frac{37747\mu}{14000x^3}-\frac{37747}{2800x}-\frac{113241}{2800\sqrt{\mu}}+\frac{113241x}{1400\mu}\biggr)
+\calo\left(Q^2\right)\,,
\end{split}
\eqlabel{1nu}
\end{equation}
where $\nu_0$ is the $\k=0$, $Q=0$ black hole warp factor \eqref{nu0}.
Given \eqref{1nu} and using 
\begin{equation}
\xp=\sqrt{\mu}-\frac 12 Q+\calo\left(Q^2\right)\,,
\eqlabel{xpq}
\end{equation}
we find\footnote{We will need the gauge potential $a(x)$ only  to order $\calo\left(Q^0\right)$.}
\begin{equation}
\begin{split}
\calc_a=&-\frac{6}{7\sqrt{\mu}}+\calo\left(Q\right)\,.
\end{split}
\eqlabel{calca}
\end{equation}

For convenience, we collect all formulas for the $\k=0$ $\a'$-corrected charged black hole 
to order $\calo\left(Q^2\right)$ in Appendix \ref{koq}.

\section{Leading $\a'$ corrections to thermodynamics of $AdS_5\times S^5$ black holes}\label{thermo}

\subsection{General framework}\label{thermof}

In this section we discuss the thermodynamics of the $\a'$ corrected black holes in asymptotic $AdS_5\times S^5$
geometry of type IIB supergravity. We assume that the following two statements are correct even in the presence of 
$\a'$ corrections:
\nxt first, for neutral black holes we have a thermodynamic relation between Helmholtz free energy density $F$, 
the energy density $E$, the entropy density $S$ and the temperature $T$
\begin{equation}
F=E-TS\,;
\eqlabel{0neutral}  
\end{equation}
second, for charged black holes we have a thermodynamic relation between Gibbs free energy density $\Omega$, 
the energy density $E$, the entropy density $S$, the chemical potential $\mu_q$ for the 
true physical charge density $\tQ$, and the temperature $T$
\begin{equation}
\Omega=E-T\ S-3\mu_q\ \tQ\,;
\eqlabel{0charged}  
\end{equation}
where the factor $3$ arises because we introduced identical chemical potential  for all three charges;
\nxt for the neutral black holes the first law of thermodynamics takes form
\begin{equation}
dF=-S\ dT\,;
\eqlabel{1neutral}
\end{equation}
while for the  charged black holes we have 
\begin{equation}
d\Omega=-S\ dT-3\tQ\ d\mu_q\,.
\eqlabel{1charged}
\end{equation}

Since we are going to rely on the validity of \eqref{0neutral}-\eqref{1charged} in black hole 
thermodynamics, it is worthwhile  recalling some nontrivial checks on the latter relations. 
In the absence of $\a'$ corrections one can independently evaluate each quantity in \eqref{0neutral}
or \eqref{0charged} and verify \eqref{1neutral} and \eqref{1charged}. Indeed, one can use holographic 
renormalization \cite{hol1,hol2,hol3}  to compute the energy density (from the one-point correlation 
function of the holographic stress-energy tensor) and  the Helmholtz ( or Gibbs) 
free energy density (as regularized Euclidean action for the background geometry). It is straightforward 
to compute Hawking temperature (surface gravity) of the black hole and the black hole entropy density
(as a quarter of the horizon area measured in Planck units).  Finally, the chemical potential can be  
determined from the value of the gauge field  at the horizon \cite{th2}, and the conserved physical charge densities  
can be determined by applying the Gauss's law.  All this was verified  in a variety of highly 
nontrivial examples\footnote{For a sample see \cite{th1,th3,th4,aby,ex1,ex2}.}.
The situation is much more complicate for black holes including  $\a'$ correction. To name one complication, 
the black hole entropy density is no longer determined simply by the area of the horizon. This was first observed 
imposing thermodynamic relation \eqref{0neutral} and \eqref{1neutral} for the $\k=0$ $\a'$-corrected 
black holes in asymptotic $AdS_5\times S^5$ type IIB supergravity \cite{gkt}. 
The authors of \cite{gkt} had to impose {\it both} \eqref{0neutral} and \eqref{1neutral} 
since they evaluated explicitly only the Helmholtz free energy density and the temperature,
while constraining the entropy density and the energy density from the consistency of the thermodynamics\footnote{
Recently is was demonstrated in \cite{g} that the entropy density computed in \cite{gkt} agrees with the Wald 
formula \cite{wald} for the black hole entropy in higher derivative gravity. Wald formula was tested as well in  
other models of higher curvature gravity \cite{rc3m2,rc3m1,rc3,rc4,rc5}.
}. Though there are few explicit checks of the $\a'$-corrected 
black hole thermodynamics, we would like to stress that validity of \eqref{0neutral}-\eqref{1charged} for 
$\a'$-corrected black holes studied here are guaranteed by the gauge theory-string theory correspondence 
\cite{m9711}. The reason for this is that the $\a'$-corrected black holes studied here 
are dual to supersymmetric $\caln=4$ $SU(n_c)$ Yang-Mills theory at finite temperature, certain 
$U(1)_R$ chemical potentials and at finite  't Hooft coupling, corresponding to finite $\a'$ corrections.

In this paper we use holographic renormalization (to leading order in $\a'$) 
to compute the free energy density and the energy density of a given black hole. 
Furthermore, we evaluate the $\a'$-corrected temperature of a black hole 
as inverse periodicity of the Euclidean time direction, requiring the absence of 
conical singularities. In the case of neutral black holes we use \eqref{0neutral} 
to compute the entropy density.  The first law of thermodynamics \eqref{1neutral} then provides 
a nontrivial consistency check on our analysis. For the charged black holes 
 we can use \eqref{0charged} to evaluate the entropy density. The remaining thermodynamic characteristics, 
\ie, the chemical potential and the physical charge, are then evaluated imposing the first law of 
thermodynamics \eqref{1charged}. In a similar way one can study thermodynamics of $\a'$-corrected 
single-charge black holes, but our approach can not be applied to black holes with two or more 
different $U(1)$ charges. 

It is convenient to introduce ``hatted'' thermodynamic  potentials as 
\begin{equation}
\Omega\equiv \frac{n_c^2}{8\pi^2}\ \hw\,,\qquad  F\equiv \frac{n_c^2}{8\pi^2}\ \hf\,,
\qquad E\equiv \frac{n_c^2}{8\pi^2}\ \he\,,\qquad 
S\equiv \frac{n_c^2}{8\pi^2}\ \hs\,,\qquad  \mu\equiv \frac{n_c^2}{8\pi^2}\ \hmu\,.
\end{equation}
Consider first computation of the Gibbs free energy density\footnote{The Gibbs free energy reduces to Helmholtz 
free energy in the absence of the chemical potentials.}. The free energy is defined as 
\begin{equation}
\frac{vol(S^3)}{\k^{3/2}}\ \Omega=T\ I_E \,,  
\end{equation}
where $I_E$ is the Euclidean action of the ``on-shell'' gravitational background. The latter is divergent and 
thus should be regularized and holographically renormalized. Define regularized Euclidean bulk effective action $S_{E,eff}^\r$
by cutting off the radial coordinate  integration
\begin{equation}
\int_{\rp}^{\infty}dr\ \to \int_{\rp}^{\r}dr
\end{equation}
in \eqref{sugrapart} and \eqref{gpart} at some UV cutoff  $r=\rho\gg \rp$. Holographically  renormalized action 
$I_E^\r$ is then obtained by supplementing $S_{E,eff}^\r$ with the Gibbons-Hawking and the boundary counterterms 
\begin{equation}
I_E^\r=S_{E,eff}^\r+I_{GH}+I_{counterterms}\,,
\eqlabel{ie0}
\end{equation}
so that no divergences arise in $\r\to\infty$ limit,
\begin{equation}
I_E=\lim_{\r\to\infty}\ I_E^\r\,.
\eqlabel{limit}
\end{equation}
The Gibbons-Hawking term is most easy to evaluate in ten dimensions. It is given by the integral over the
nine-dimensional boundary $\del\calm_{10}$ located at $r=\r$ 
\begin{equation}
I_{GH}=-\frac{1}{8\pi G_{10}}\int_{\del\calm_{10}, r=\r} d^9\xi \sqrt{h_E}\nabla_\mu n^\mu\,,
\eqlabel{GH}
\end{equation}
where $h_E$ is induced metric on $\del\calm_{10}$ and $n^\mu=D_3^{-1}\dd^\mu_r$ is a unit outward normal to this 
boundary. For the boundary counterterms we would like to use results of \cite{th3,th4}. For this reason we express 
the counterterm action $I_{counterterm}$ as 
\begin{equation}
I_{counterterms}=\frac{1}{4\pi G_5} \int_{\del\calm_5, r=\r}d^4\xi \biggr(\a_1\sqrt{H_E}+\a_2 R_4\sqrt{H_E}\biggr)\,,
\eqlabel{bcounter}
\end{equation}
where $\del\calm_5$ is a Kaluza-Klein reduction of $\del\calm_{10}$, the five-dimensional Newton's constant 
is $G_5=\frac{G_{10}}{\pi^3}$, and $R_4$ is a Ricci scalar evaluated on the  metric $H_{E,\mu\nu}$ which is induced 
on the $\del\calm_5$ boundary from the five-dimensional Kaluza-Klein reduced metric  
\begin{equation}
{G_{E,\mu\nu}}\ d\xi^{\mu}d\xi^{\nu}=D_4^{10/3}\ \biggl( D_1^2 dt^2+ D_2^2\ \frac{1}{\k}\left(S^3\right)^2+D_3^2dr^2 \biggr) \,. 
\eqlabel{metric5}
\end{equation}
In the supergravity approximation (see \cite{th3} where the same notations have been used)
\begin{equation}
\a_1=\frac 32\,,\qquad \a_2=\frac 18\,.
\eqlabel{alphas}
\end{equation}
Though in principle $\a_i$ in \eqref{alphas} might receive $\a'$-corrections, for the class of black holes discussed here, 
this does not happen. The reason is that $\a'$-corrections to the charged black hole metric 
vanish sufficiently fast at asymptotic infinity and does not affect the divergences of  $S_{E,eff}^\r$ as $\r\to\infty$.

Since each  contribution of $I_E^\r$ is evaluated with the  $\a'$-corrected on-shell  metric \eqref{1order}, we can naturally 
split (to leading order in $\g$ \eqref{defg})
\begin{equation}
\begin{split}
&I_E^\rho\equiv I_{0,E}^\r+\g\ \dd I_E\\
&=\biggl(S_{0,E,eff}^\r+I_{0,GH}+I_{0,counterterms}\biggr)+\g\ \biggl(\dd I_{SUGRA}+\dd I_{W}+\dd 
I_{GH}+\dd I_{counterterms}\biggr)\,,
\end{split}
\eqlabel{split}
\end{equation} 
where the subscript $ _0$ indicates purely supergravity contributions, while $\dd\cdots$ denotes $\a'$-corrections. 
Notice that 
$$\frac{T}{vol(S^3)/\k}\ \lim_{r\to\infty} I_{0,E}^\r
$$
gives precisely the Gibbs free energy density in \eqref{charged}.

The $\dd I_{SUGRA}$ contribution comes the supergravity part \eqref{sugrapart} of the effective action \eqref{seffsugra}.
Due to the diffeomorphism invariance, much like $S_{0,E,eff}$, $\dd I_{SUGRA}$ can be written as a total derivative 
involving the deformations $\{A,B,\nu,a\}$ in \eqref{1order}, and thus 
receives contributions only from the regularization $r=\r$ boundary and the horizon $r=\rp$.  Explicitly,
\begin{equation}
\begin{split}
\dd I_{SUGRA}=&\frac{vol(S^3)}{T\k^{3/2}}\ \frac{\pi^3}{16\pi G_{10}}\ \biggl(
((4 Q \mu+4 \k Q^2)+(-4 \mu-8 \k Q) x+4 \k x^2+4 x^3)\ B'\\
&-\frac {2}{3x} 
((-24 Q \mu-24 \k Q^2) x+(48 \k Q+24 \mu) x^2-24 \k x^3-24 x^4)\ A'\\
&-\frac 2x ((Q \mu+\k Q^2)+(\mu+2 \k Q) x-3 \k x^2-5 x^3) B\\
&-\frac 2x ((Q \mu+\k Q^2)+(-4 \k Q-2 \mu) x+3 \k x^2+4 x^3) A\\
&-\frac{20}{3} ((Q \mu+\k Q^2)+(-\mu-2 \k Q) x+\k x^2+x^3)\ \nu'\\
&+\frac{2 Q (\mu+\k Q)}{x}\  a
\biggr)\ \bigg|_{x=\xp}^{x=\r^2+Q}\,,
\end{split}
\eqlabel{disugra}
\end{equation}
where we used the fact that Euclidean time direction is periodic with periodicity $\frac 1T$.  
Notice that the coefficient in front of $\nu'$ in \eqref{disugra} vanishes\footnote{This term does not contribute when evaluated 
at the  boundary 
$x=\r^2+Q$  either as $\nu(x)$ vanishes fast enough, see \eqref{nuass}.} for $x=\xp$, thus even though for $\k\ne 0$ 
neutral black holes  ($Q=0$) we can not find the analytic solution for $\nu(x)$, we can explicitly compute the 
$\a'$ corrections contributed by \eqref{disugra}. For the charged black holes we need to know the value of the 
gauge potential $a$ at the horizon, which through $\calc_a$ \eqref{b2ca}  ( or \eqref{b2ca0} ) requires the knowledge of 
$\nu(x)$. In fact, above is the only place where the $\a'$-corrected thermodynamics of charged black holes in 
asymptotic $AdS_5\times S^5$ geometry is sensitive to the five-sphere warp factor $\nu$.     

The $\dd I_W$ is most easy to evaluate as a (convergent) bulk integral, much like it was done in \cite{gkt}
\begin{equation}
\begin{split}
\dd I_{W}=&-\frac{vol(S^3)}{T\k^{3/2}}\ \frac{\pi^3}{16\pi G_{10}}\ \int_{\rp}^{\infty}dr\ D_1^{(0)} \left[D_2^{(0)}\right]^3\ 
D_3^{(0)} \left[D_5^{(0)}\right]^5 \
\ W\left[D_i^{0},Da^{(0)}\right]\,.
\end{split}
\eqlabel{diddw}
\end{equation}
Notice that the integrand in \eqref{diddw} is evaluated with the $a'$-uncorrected metric \eqref{zeroorder}.

Finally, $\dd I_{GH}$ and $\dd I_{counterterms}$ are simply the contributions $\propto \g$ coming from 
\eqref{GH} and \eqref{bcounter} (with \eqref{alphas}) when evaluated with $\a'$-corrected metric \eqref{1order}.

In addition to the Gibbs free energy density we need to evaluate the energy density, $E$. 
We find\footnote{We follow here the  presentation of  \cite{th3}.} 
\begin{equation}
\frac{vol(S^3)}{\k^{3/2}}\ E=\int_{\Sigma}d^3\xi\ \sqrt{\sigma} N_{\Sigma} \epsilon\,,
\eqlabel{massdef}
\end{equation}
where $\Sigma\equiv S^3$ is a spacelike hypersurface in $\del\calm_5$
with a timelike unit normal $u^\mu$,
$N_{\Sigma}$ is the norm of the timelike Killing vector 
in \eqref{metric5}, ${\sigma}$ is the determinant of the induced metric 
on $\Sigma$, and $\epsilon$ is the proper energy density 
\begin{equation}
\epsilon=u^\mu u^\nu T_{\mu\nu}\,.
\eqlabel{epdef}
\end{equation}
The quasilocal stress tensor $T_{\mu\nu}$ for our background
is obtained from the variation of the Kaluza-Klein reduced Gibbons-Hawking term \eqref{GH} and the 
boundary counterterms \eqref{bcounter}\footnote{There is no contribution from the bulk effective action \eqref{seffsugra}.}
with respect to the boundary metric $\delta H_{\mu\nu}$
\begin{equation}
T^{\mu\nu}=\frac{2}{\sqrt{-H}}\ \frac{\delta }{\delta H_{\mu\nu}}\ \biggl[I_{GH}+I_{counterterms}\biggr]\,.
\eqlabel{qlst}
\end{equation}  
Explicit computation yields 
\begin{equation}
T^{\mu\nu}=\frac{1}{8\pi G_5}\biggl[
-\Theta^{\mu\nu}+\Theta H^{\mu\nu}-2\a_1 H^{\mu\nu}+4\a_2 \left(
R_4^{\mu\nu}-\ft 12 R_4 H^{\mu\nu}\right)\biggr]\,,
\eqlabel{tfin}
\end{equation}
where 
\begin{equation}
\Theta^{\mu\nu}=\ft 12 \left(\nabla^\mu n^\nu+
\nabla^\nu n^\mu\right),\qquad \Theta=\tr \Theta^{\mu\nu}\,.
\eqlabel{thdef}
\end{equation}

\subsection{Thermodynamics of neutral black holes}
We use the framework detailed  in Section \ref{thermof} to study the thermodynamics of $\a'$-corrected neutral black holes
in asymptotic $AdS_5\times S^5$ type IIB supergravity with the spatial horizon curvature $\k=\{0,1\}$. 
The relevant formulas for the $\a'$-corrected geometry are given in Section \ref{neutralf}.  
The only other information we need is the value of the quartic Weyl invariant $W$ \eqref{rrrr} evaluated in the 
supergravity approximation. We find 
\begin{equation}
W=\frac{180\mu^4}{r^{16}}\,.
\eqlabel{wneutral}
\end{equation}

Recalling that the horizon  location is given by the largest positive root of \eqref{horizon},  
\begin{equation}
0=\xp^2+\xp\k-\mu\,,
\eqlabel{xpneutral}
\end{equation}
explicit evaluation of various contribution in \eqref{split} yields  
\begin{equation}
\begin{split}
\dd I_{SUGRA}=\frac{vol(S^3)}{T\k^{3/2}}\ \frac{\pi^3}{16\pi G_{10}}\ \times\ \biggl\{
-\frac{15\mu^3}{7\xp^6}\biggl(31\mu-24\k\ \xp\biggr)
\biggr\}\,,
\end{split}
\eqlabel{bulkcc}
\end{equation}
\begin{equation}
\begin{split}
\dd I_{W}=\frac{vol(S^3)}{T\k^{3/2}}\ \frac{\pi^3}{16\pi G_{10}}\ \times\ \biggl\{
-\frac{15\mu^4}{\xp^6}
\biggr\}\,,
\end{split}
\eqlabel{Wcc}
\end{equation}
\begin{equation}
\begin{split}
\dd I_{GH}=\frac{vol(S^3)}{T\k^{3/2}}\ \frac{\pi^3}{16\pi G_{10}}\ \times\ \biggl\{
\frac{20\mu^3}{7\xp^6}\biggl(57\mu-32\k\ \xp\biggr)
\biggr\}\,,
\end{split}
\eqlabel{GHcc}
\end{equation}
\begin{equation}
\begin{split}
\dd I_{counteterms}=\frac{vol(S^3)}{T\k^{3/2}}\ \frac{\pi^3}{16\pi G_{10}}\ \times\ \biggl\{
-\frac{15\mu^3}{7\xp^6} (57\mu-32\k\ \xp)
\biggr\}\,,
\end{split}
\eqlabel{countercc}
\end{equation}
so that for the Helmholtz free energy density we find
\begin{equation}
\hf=-\mu+\frac 34\k^2+2\k\xp+\g\times\biggl\{-\frac {5\mu^3}{7\xp^6} \biggl(57\mu-40\k\xp\biggr)\biggr\}\,.
\eqlabel{fneutral}
\end{equation}
For the energy density $E$ \eqref{massdef} we find
\begin{equation}
\he=3\mu+\frac 34\k^2+\g\times\biggl\{\frac{3}{28\xp^2}\biggl(
(500\k^3\xp+500\k^4)+(2140\k^2+1640\k\xp)\mu+1140\mu^2\biggr)\biggr\}\,.
\eqlabel{massneutral}
\end{equation}
Finally, the Hawking temperature is 
\begin{equation}
\pi T=\frac{\mu}{2\xp^{3/2}}+\frac{\sqrt{\xp}}{2}+\g\times\biggl\{
\frac{5}{7\xp^{7/2}}\biggl(-\k^4-\k^3\xp+(8\k\xp+7\k^2)\mu+9\mu^2\biggr)\biggr\}\,.
\eqlabel{etaneutr}
\end{equation}

Using the basic thermodynamic relation \eqref{0neutral} we find the entropy density
\begin{equation}
\hs=4\pi\xp^{3/2}+\g\times\biggl\{\frac{960\pi}{7\xp^{3/2}}\biggl(\k^2\xp+\k^3+(\xp+2\k)\mu\biggr)\biggr\}\,.
\eqlabel{sneutral}
\end{equation}
It is straightforward to verify that for arbitrary spatial horizon curvature $\k$ and to leading  order in $\g$, 
the first law of thermodynamics \eqref{1neutral} is automatically satisfied.

We have now all the necessary information to compute the $\a'$ corrections to the Hawking-Page phase transition 
temperature. Since global $AdS_5\times S^5$ geometry does not receive $\a'$ corrections to leading order 
(see Section \ref{neutralf}), the transition temperature $T_{HP}$ is determined from the condition
\begin{equation}
\hf(\mu)=\he(\mu=0,\g=0)\qquad \Rightarrow\qquad \pi T_{HP}=\frac 32\ \k^{1/2}\ \biggl\{1-40\g\biggr\}\,,
\eqlabel{eqlabel{tgpa}}
\end{equation}
where $\hf$ and $\he$ are given in \eqref{fneutral} and \eqref{massneutral} correspondingly.

\subsection{Thermodynamics of charge black branes at $\mu_q\ll T$}
We use the framework detailed  in Section \ref{thermof} to study the thermodynamics of $\a'$-corrected charged  black holes
in asymptotic $AdS_5\times S^5$ type IIB supergravity with the spatial horizon curvature $\k=0$ and in the high temperature regime,
\ie, $T\gg \mu_q$. 
The relevant formulas for the $\a'$-corrected geometry are given in Section \ref{chargedf} (or  Appendix C).  
The only other information we need is the value of the quartic Weyl invariant $W$ \eqref{rrrr} evaluated in the 
supergravity approximation. We find\footnote{All expressions in this section are to leading order in $Q$.} 
\begin{equation}
\begin{split}
W=&\frac{180\mu^4}{x^8}-\frac{240\mu^3 Q}{x^{9}}\biggl(4 \mu+ x^2\biggr)\,.
\end{split}
\eqlabel{wcharged}
\end{equation}

Recalling that the horizon  location is given by (see \eqref{eqw1}),  
\begin{equation}
\xp=\sqrt{\mu}-\frac 12 Q
\eqlabel{xpcharged}
\end{equation}
explicit evaluation of various contribution in \eqref{split} yields  
\begin{equation}
\begin{split}
\dd I_{SUGRA}=\frac{vol(S^3)}{T\k^{3/2}}\ \frac{\pi^3}{16\pi G_{10}}\ \times\ \biggl\{
-\frac{465}{7}\ \mu+Q\left(
\frac{4245}{7}+\frac{3853}{18}\ \sqrt{\mu}
\right)
\biggr\}\,,
\end{split}
\eqlabel{bulkccc}
\end{equation}
\begin{equation}
\begin{split}
\dd I_{W}=\frac{vol(S^3)}{T\k^{3/2}}\ \frac{\pi^3}{16\pi G_{10}}\ \times\ \biggl\{
-15\mu+\frac{333}{7}\ \sqrt{\mu}Q 
\biggr\}\,,
\end{split}
\eqlabel{Wccc}
\end{equation}
\begin{equation}
\begin{split}
\dd I_{GH}=\frac{vol(S^3)}{T\k^{3/2}}\ \frac{\pi^3}{16\pi G_{10}}\ \times\ \biggl\{
\frac{1140}{7}\ \mu-\frac{32749}{63}\sqrt{\mu} Q
\biggr\}\,,
\end{split}
\eqlabel{GHccc}
\end{equation}
\begin{equation}
\begin{split}
\dd I_{counteterms}=\frac{vol(S^3)}{T\k^{3/2}}\ \frac{\pi^3}{16\pi G_{10}}\ \times\ \biggl\{
-\frac{855}{7}\mu+\frac{32749}{84}\sqrt{\mu} Q
\biggr\}\,,
\end{split}
\eqlabel{counterccc}
\end{equation}
so that for the Gibbs free energy density we find
\begin{equation}
\hw=-\mu+\g\times \biggl\{-\frac{285}{7}\mu+\left(\frac{33181}{252}\sqrt{\mu}+\frac{4245}{7}\right)Q\biggr\}\,.
\eqlabel{gcharged}
\end{equation}
For the energy density $E$ \eqref{massdef} we find
\begin{equation}
\he=3\mu+\g\times \biggl\{\frac{855}{7}\mu-\frac{32749}{84}\sqrt{\mu}Q\biggr\}\,.
\eqlabel{masscharged}
\end{equation}
Finally, the Hawking temperature is 
\begin{equation}
\pi T={\mu}^{1/4}-\frac {3}{4\mu^{1/4}}\ Q+\g\times \biggl\{\frac{45}{7}{\mu}^{1/4}+\frac{2243}{1008\mu^{1/4}}\ Q\biggr\}\,.
\eqlabel{etacharged}
\end{equation}

As explained in Section \ref{thermof}, we use the basic thermodynamic relation \eqref{0charged} and the first law of thermodynamics 
\eqref{1charged}
to determine the remaining thermodynamic potentials, \ie, the entropy density $\hs$, the chemical potential 
$\hmu$ and the physical charge density $\tQ$. 
Specifically, we parameterize the chemical potential $\hmu$ and the physical 
charge $\tQ$ (to leading order in $\g$ and to leading order in $Q$) as 
\begin{equation}
\hmu=2Q^{1/2}\biggl(1+\g\ \dd\mu_q(\mu)\biggr)\,, \qquad \tQ=Q^{1/2}\biggl(\sqrt{\mu}+\g\ \dd q(\mu)\biggr)\,.
\eqlabel{muQpar}
\end{equation}
Next, we evaluate the entropy density from \eqref{0charged}, and impose the first law of thermodynamics \eqref{1charged}
\begin{equation}
\begin{split}
0=&d\hw+\hs\ dT+3\tQ\ d\hmu=\g\times \biggl\{
\frac{264}{7}\mu^{1/2}+\frac{4245}{7}+3\ \dd q(\mu)+3\sqrt{\mu}\ \dd\mu_q(\mu)
\biggr\}\ d(Q)\\
&+\g\times \biggl\{
-\frac{4245}{28\mu} -\frac{123}{7\sqrt{\mu}}-\frac {3}{2\mu}\ \dd q(\mu)+6\sqrt{\mu}\ \frac{d}{d\mu} \dd \mu_q(\mu)-\frac {3}{2\sqrt{\mu}}
\dd \mu_q(\mu)
\biggr\}\ Q d(\mu)\,.
\end{split}
\eqlabel{1lawc}
\end{equation}
From \eqref{1lawc} we find
\begin{equation}
\dd\mu_q(\mu)=-\frac{3}{14}\ \ln\frac{\mu}{\pi^4\epsilon^4}+\frac{1415}{28\sqrt{\mu}}-\frac{89}{7}\,,\qquad 
\dd q(\mu)=-\frac{7075}{28}+\frac 17\sqrt{\mu}+\frac{3}{14}\sqrt{\mu}\ \ln\frac{\mu}{\pi^4\epsilon^4}\,.
\eqlabel{solve1law}
\end{equation}
where $\epsilon$ is an arbitrary energy scale, arising as an integration constant in solving \eqref{1lawc}.
Though $\hmu$ and $\tQ$ separately depend on $\epsilon$, their  product $\hmu \tQ$, and thus 
via \eqref{0charged} the entropy density $\hs$, is $\epsilon$ independent. It would be interesting to
repeat computation of \cite{g} and compare the entropy density obtained here from the thermodynamics 
with the Wald formula \cite{wald}.  
We comment on the physical meaning of the energy scale $\epsilon$ in the conclusion section. 

In the rest of this section we express all thermodynamic properties of the $\a'$-corrected charged black holes 
discussed here in 
terms of the  temperature $T$ and the physical charge chemical potential $\mu_q$. We find\footnote{Correct dimensionality 
can be restored with $L$, with was set to unity.}
\begin{equation}
\begin{split}
\hw=&-\pi^4 T^4\biggl\{\ \biggl(1+15\times \g+\calo(\g^2)\biggr)\\
&+\frac{3\mu_q^2}{4\pi^2T^2}\left(1-\g\times \left(\frac{4245}{14\pi^2 T^2}+\frac{12}{7}\ln \frac{\epsilon}{T}\right)
+\calo(\g^2)
\right)\\
&+\calo\left(\frac{\mu_q^4}{T^4}\right)\biggl(1+ \calo(\g)\biggr)\ \biggr\}\,,
\end{split}
\eqlabel{endgibbs}
\end{equation} 
\begin{equation}
\begin{split}
\he=&3\pi^4 T^4\biggl\{\ \biggl(1+15\times \g+\calo(\g^2)\biggr)\\
&+\frac{3\mu_q^2}{4\pi^2T^2}\left(1+\g\times \left(\frac 47-\frac{1415}{14\pi^2 T^2}-\frac{12}{7}\ln \frac{\epsilon}{T}\right)
+\calo(\g^2)
\right)\\
&+\calo\left(\frac{\mu_q^4}{T^4}\right)\biggl(1+ \calo(\g)\biggr)\ \biggr\}\,,
\end{split}
\eqlabel{endenergy}
\end{equation} 
\begin{equation}
\begin{split}
\hs=&3\pi^4 T^3\biggl\{\ \biggl(1+15\times \g+\calo(\g^2)\biggr)\\
&+\frac{3\mu_q^2}{8\pi^2T^2}\left(1+\g\times \left(\frac 67-\frac{12}{7}\ln \frac{\epsilon}{T}\right)
+\calo(\g^2)
\right)\\
&+\calo\left(\frac{\mu_q^4}{T^4}\right)\biggl(1+\calo(\g)\biggr)\ \biggr\}\,,
\end{split}
\eqlabel{endentropy}
\end{equation} 
\begin{equation}
\begin{split}
\tQ=&\frac 12 \mu_q \pi^2 T^2\biggl\{\ \biggl(1-\g\times \left(\frac{4245}{14\pi^2 T^2}+\frac{12}{7}\ln \frac{\epsilon}{T}\right)+\calo(\g^2)\biggr)+\calo\left(\frac{\mu_q^2}{T^2}\right)\biggl(1+ \calo(\g)\biggr)\ \biggr\}\,.
\end{split}
\eqlabel{endqf}
\end{equation} 
Notice that in the absence of chemical potential, $\mu_q=0$, we reproduce results of \cite{gkt,pt}.

\section{Conclusion}
In this paper we studied leading $a'$ corrections to charged near-extremal black holes in asymptotic 
$AdS_5\times S^5$ type IIB supergravity. We  used $\a'$-corrected holographic renormalization
to evaluate the free energy density and the ADM  mass density  of the black holes under the assumption 
that the five-form does not receive $\a'$ corrections. We extracted the remaining 
thermodynamic  properties imposing the basic thermodynamic relations (\eqref{0neutral} or \eqref{0charged})
along with the first law of thermodynamics.  

In the case of neutral black holes with spatial horizon curvature $\k$ we showed that given the Helmholtz 
free energy and the energy density, the entropy density 
obtained from the basic thermodynamic relation \eqref{0neutral} automatically satisfies the first law of thermodynamics. 
We evaluated leading $\a'$ corrections to the Hawking-Page phase transition temperature. Hawking-Page phase transition 
is a holographic dual \cite{w98} to a confinement-deconfinement phase transition in $\caln=4$ $SU(n_c)$ SYM 
compactified on $S^3$, which was extensively studied recently at weak t' Hooft coupling in \cite{ah1,ah2}.  
   
Our computational approach allows to study either black holes with a single nonzero charge under the $[U(1)_R]^3$ 
symmetry or  black holes with all identical $U(1)_R$ symmetry charges. In this paper we discussed technically 
less complicated case of identical charges\footnote{The thermodynamics of 
single-charge nonextremal black holes in asymptotic $AdS_5\times S^5$ type IIB supergravity 
will be discussed elsewhere.}.  Apart from the $S^5$ sphere warp factor, we analytically 
solved for the $\a'$-corrected nonextremal black hole geometry with the spatial horizon curvature $\k=\{0,1\}$.
Unlike the neutral black holes, the knowledge of the latter warp factor is vital for understanding $\a'$ corrected  
thermodynamic properties of charged black holes. We solved for the $S^5$ warp factor of the $\k=0$ 
charged black hole analytically in the 
high temperature regime, when the temperature is much larger than the chemical potential. 
Rather interestingly, we found that in the high  temperature regime, the Gibbs free energy density 
of  these black holes receives 
correction \eqref{endgibbs}
\begin{equation}
\dd \Omega \propto -(\a')^3\ T^2 \mu_q^2\ \ln \frac {T}{\epsilon} \,,
\eqlabel{lncorr}
\end{equation}
where $\epsilon$ is an arbitrary scale. From the dual gauge theory perspective it appears that the 
Gibbs free energy of the $\caln=4$ SYM at finite temperature, chemical potential and a finite (but large) 't Hooft coupling
depends on arbitrary energy scale $\epsilon$. How  is it possible?

To understand this puzzle, consider four dimensional, non-supersymmetric 
QCD with quarks, at finite temperature. To two-loop order using hard-thermal-loop perturbation theory
the Helmholtz free energy density (or minus the pressure) is given by \cite{aps}
\begin{equation}
F_{QCD}=-\frac{8\pi^2}{45}T^4\ \biggl[\calf_0+\calf_2\ \frac{\a_s}{\pi}+\calf_3\ \left(\frac{\a_s}{\pi}\right)^{3/2}
+\calf_4\ \left(\frac{\a_s}{\pi}\right)^{2}+\calf_5\ \left(\frac{\a_s}{\pi}\right)^{5/2}
+\calo\left(\a_s^3\ln \a_s\right)\biggr]
\eqlabel{qcd}    
\end{equation}
where  $\calf_4$ and $\calf_5$ contain an explicit $\ln \frac{\epsilon}{T}$ dependence on the 
renormalization scale $\epsilon$. 
There is no contradiction here, because taking into account the running coupling constant $\a_s=\a_s(\epsilon)$
one finds\footnote{Renormalization group resummation of the free energy of hot QCD 
was discussed in \cite{gerry}.} that 
\begin{equation}
\epsilon \frac{d}{d\epsilon }\ F_{QCD}=\calo\left(\a_s^3\ln \a_s\right)\,.
\end{equation}
Turning things around, appearance of the $\ln \frac{\epsilon}{T}$  signals the $\epsilon$-dependence 
( the familiar perturbative running) of the gauge coupling $\a_s$. Thus, we are tempted to conjecture 
that appearance of an arbitrary energy scale $\epsilon$ in \eqref{lncorr} signals the scale dependence, 
``running'' of the inverse 't Hooft coupling $\lambda^{-1}=1/(g_{YM}^2 n_c)\propto \g^{3/2}=\g^{3/2}(\epsilon)$
in $\caln=4$ SYM. If so, requiring that $\hw$ in \eqref{endgibbs} is a renormalization group (RG) invariant to order computed,
\ie, 
\begin{equation}
\epsilon \frac{d}{d\epsilon }\ \hw=0+\calo(\g^2)+\calo\left(\frac{\mu_q^4}{T^4}\right)\biggl(1+\calo(\g)\biggr)\,,
\eqlabel{rgiggs}
\end{equation}
 we find 
\begin{equation}
\epsilon \frac{d}{d\epsilon }\ \g=\frac{3\mu_q^2}{35\pi^2 T^2}\ \g\,.
\eqlabel{betag}
\end{equation}
 The RG running \eqref{betag} is to be understood as running of the effective $\caln=4$ coupling  at energy 
scales below $T$, which is the highest scale in the problem. We do not expect running at energies much higher than the 
temperature scale. In this respect the situation is similar to a mass deformation of the $\caln=4$ SYM: {\it above} the mass 
scale, the SYM coupling is constant, while the effective Wilsonian coupling runs {\it below} the mass scale 
since some (or all) of the scalars and fermions of the $\caln=4$ supersymmetric Yang-Mills are integrated out. 

Finally, if there is running of the effective $\caln=4$ SYM coupling at strong coupling (and nonzero chemical potential) 
it is natural to conjecture that SYM coupling would also run at weak 't Hooft coupling
and temperatures much larger than the chemical potential. The structure of the perturbative 
(in the 't Hooft coupling $\lambda$) corrections to a hot QCD pressure \cite{aps} and that of the 
$\caln=4$ gauge theory plasma \cite{py1,py2,py3} is similar to order $\lambda^{3/2}$. Unfortunately, 
there is no computation in thermal $\caln=4$ available to order $\lambda^2$, which is the first order 
in perturbation theory where $\ln \frac{\epsilon}{T}$ correction appears in QCD \eqref{qcd}. Actually, 
we expect that the $\caln=4$ gauge coupling will run only when the chemical potential is nonzero. 
But again, currently, the computations with nonzero chemical potential 
are available only to order $\lambda^{3/2}$ \cite{larry}.

\section*{Acknowledgments}
I would like to thanks  Philip Argyres, Stan Deakin, Jaume Gomis,  Gerry Mckeon, Rob Myers and  Volodya Miransky
for valuable discussions. I would like to thank  
Banff International Research Station,  Michigan Center for Theoretical Physics and  University of Cincinnati for hospitality 
where part of this work was done.
Research at Perimeter Institute is supported in part by funds from NSERC of
Canada. I gratefully   acknowledge  support by  NSERC Discovery
grant.

\section*{Appendix}
\appendix

\section{Data for charged $\k=1$ black holes}\label{k1}
Inhomogeneous parts of equations \eqref{Ameq1}-\eqref{cmeq1}:
\begin{equation}
\begin{split}
J_A=&-\frac{1}{12x^{11}} \biggl(-160 Q (Q+\mu) (2 Q+\mu) x^4+(-23040 Q^3+912 \mu^2 Q^2+912 Q^4\\
&-34560 \mu Q^2-4320 \mu^3
-20160 \mu^2 Q+1824 Q^3 \mu) x^3+ 6 Q (Q+\mu) (5359 \mu^2\\
&+18440 Q \mu+18440 Q^2) x^2-84196 Q^2 (Q+\mu)^2 (2 Q+\mu) x+82707 Q^3 (Q+\mu)^3\biggr)
\end{split}
\eqlabel{sourceA1}
\end{equation}
\begin{equation}
\begin{split}
J_B=&\frac{1}{8 x^{11} (x^2-2 x Q+Q^2+x^3-\mu x+Q \mu)} \biggl(704 Q (Q+\mu) (2 Q+\mu) x^7+(-4608 Q^3 \mu\\
&-2304 Q^4-84672 \mu Q^2-56448 Q^3-11520 \mu^3-51264 \mu^2 Q-2304 \mu^2 Q^2) x^6\\
&+( 553536 Q^3 \mu+360304 \mu^2 Q^2-50560 \mu^2 Q-56320 Q^3+83536 \mu^3 Q-11200 \mu^3\\
&-84480 \mu Q^2+276768 Q^4) x^5+( -428544 Q^5+11160 \mu^4-1071360 \mu Q^4\\
&+155468 \mu^3 Q-214272 \mu^3 Q^2+389600 Q^4+779200 Q^3 \mu+545068 \mu^2 Q^2\\
&-857088 \mu^2 Q^3) x^4+ 8 Q (Q+\mu) (26678 Q^4+53356 Q^3 \mu-129602 Q^3+26678 \mu^2 Q^2\\
&-194403 \mu Q^2-88139 \mu^2 Q-11669 \mu^3) x^3+ 2 Q^2 (Q+\mu)^2 (147484 \mu^2+670389 Q \mu\\
&+670389 Q^2) x^2-423932 Q^3 (Q+\mu)^3 (2 Q+\mu) x+210625 Q^4 (Q+\mu)^4\biggr)
\end{split}
\eqlabel{sourceB1}
\end{equation}
\begin{equation}
\begin{split}
J_{\nu}=&\frac{1}{320 x^{11} (x^2-2 x Q+Q^2+x^3-\mu x+Q \mu)} \biggl(47488 Q (Q+\mu) (2 Q+\mu) x^7 \\
&-64 Q (Q+\mu) (-1624 \mu+2511 Q \mu-3248 Q+2511 Q^2) x^6+( -982784 \mu^2 Q^2\\
&-850240 Q^4-15360 \mu^2 Q-15360 \mu^3+30720 Q^3-1700480 Q^3 \mu+46080 \mu Q^2\\
&-132544 \mu^3 Q) x^5+( 31320 \mu^4-63360 Q^3 \mu+561344 \mu^3 Q^2+2806720 \mu Q^4\\
&+1122688 Q^5-31680 Q^4+2245376 \mu^2 Q^3+169728 \mu^3 Q+138048 \mu^2 Q^2) x^4 \\
&-16 Q (Q+\mu) (30081 Q^4+60162 Q^3 \mu+17464 Q^3+30081 \mu^2 Q^2+26196 \mu Q^2\\
&+28376 \mu^2 Q+9822 \mu^3) x^3+ 4 Q^2 (Q+\mu)^2 (73187 \mu^2+188384 Q \mu+188384 Q^2) x^2 \\
&-349232 Q^3 (Q+\mu)^3 (2 Q+\mu) x+225375 Q^4 (Q+\mu)^4\biggr)
\end{split}
\eqlabel{sourcenu1}
\end{equation}
\begin{equation}
\begin{split}
J_{a}=&\frac{1}{2x^{11}} \biggl(-576 Q (Q+\mu) x^6+(3840 \mu^2+9984 Q^2+9984 Q \mu) x^5+( -33440 Q^3\\
&+13440 Q^2+13440 Q \mu-50160 \mu Q^2+4800 \mu^2-16720 \mu^2 Q) x^4+( 25464 Q^4\\
&-104688 \mu Q^2-44976 \mu^2 Q+25464 \mu^2 Q^2-5040 \mu^3-69792 Q^3+50928 Q^3 \mu) x^3\\
&+ 14 Q (Q+\mu) (1909 \mu^2+9372 Q \mu+9372 Q^2) x^2 -53200 Q^2 (2 Q+\mu) (Q+\mu)^2 x\\
&+31473 Q^3 (Q+\mu)^3\biggr)
\end{split}
\eqlabel{sourcena1}
\end{equation}
\begin{equation}
\begin{split}
J_{const}=&-\frac{1}{24 x^{10} }\biggl(1696 Q (Q+\mu) (2 Q+\mu) x^7 -128 Q (Q+\mu) (31 Q \mu-22 \mu+31 Q^2\\
&-44 Q) x^6+(-31352 \mu^2 Q^2-11520 \mu Q^2-7680 Q^3-51392 Q^3 \mu-5656 \mu^3 Q\\
&-7680 \mu^2 Q-25696 Q^4-1920 \mu^3) x^5+( 81192 \mu^2 Q^2+27816 \mu^3 Q+101600 \mu Q^4\\
&+81280 \mu^2 Q^3+106752 Q^3 \mu+3240 \mu^4+20320 \mu^3 Q^2+53376 Q^4+40640 Q^5) x^4\\
& -4 Q (Q+\mu) (5211 Q^4+10422 Q^3 \mu+37072 Q^3+5211 \mu^2 Q^2+55608 \mu Q^2\\
&+28756 \mu^2 Q+5110 \mu^3) x^3+4 Q^2 (51173 Q^2+51173 Q \mu+13461 \mu^2) (Q+\mu)^2 x^2 \\
&-69972 Q^3 (2 Q+\mu) (Q+\mu)^3 x+37841 Q^4 (Q+\mu)^4\biggr)
\end{split}
\eqlabel{sourcec1}
\end{equation}

Inhomogeneous part of equation \eqref{final1}:
\begin{equation}
\begin{split}
\cali_{\nu}=&-\frac{1}{ 320 x^{11} (x^2-2 x Q+Q^2+x^3-\mu x+Q \mu)} 
\biggl(-33920 Q (2 Q+\mu) (Q+\mu) x^7\\
&+64 Q (Q+\mu) (-1272 \mu+1919 Q \mu+1919 Q^2-2544 Q) x^6+ 64 Q (Q+\mu) (1844 \mu^2\\
&+11321 Q \mu-720 \mu+11321 Q^2-1440 Q) x^5+( -2528960 \mu Q^4+83520 \mu^3 Q\\
&-1011584 Q^5-5400 \mu^4+544704 Q^4-505792 \mu^3 Q^2+1089408 Q^3 \mu+628224 \mu^2 Q^2\\
&-2023168 \mu^2 Q^3) x^4 -16 Q (Q+\mu) (2078 \mu^3+44128 \mu^2 Q-28147 \mu^2 Q^2-56294 Q^3 \mu\\
&+119916 \mu Q^2-28147 Q^4+79944 Q^3) x^3+ 4 Q^2 (65045 \mu^2+370952 Q \mu\\
&+370952 Q^2) (Q+\mu)^2 x^2 -423344 Q^3 (Q+\mu)^3 (2 Q+\mu) x+189257 Q^4 (Q+\mu)^4\biggr)
\end{split}
\eqlabel{i1}
\end{equation}

Inhomogeneous part of equation \eqref{dil1}:
\begin{equation}
\begin{split}
J_{\phi}=&-\frac{1}{16 x^{11} (-\mu x+x^3+Q^2-2 Q x+x^2+Q \mu)} \biggl(576 Q^2 (Q+\mu)^2 x^6\\
& -288 Q (Q+\mu) (5 \mu^2+14 Q \mu+14 Q^2) x^5+( 1080 \mu^4+11520 \mu^2 Q^3+14400 Q^4 \mu\\
&+2880 Q^2 \mu^3+7488 Q^4+14976 \mu Q^3+13248 Q^2 \mu^2+5760 Q \mu^3+5760 Q^5) x^4 \\
&-576 Q (Q+\mu) (3 Q^4+52 Q^3+6 \mu Q^3+3 Q^2 \mu^2+78 \mu Q^2+46 Q \mu^2+10 \mu^3) x^3\\
&+36 Q^2 (1304 Q^2+1304 Q \mu+383 \mu^2) (Q+\mu)^2 x^2 -16992 Q^3 (2 Q+\mu) (Q+\mu)^3 x\\
&+9531 Q^4 (Q+\mu)^4\biggr)
\end{split}
\eqlabel{jphi1}
\end{equation}

\section{Data for charged $\k=0$ black holes}\label{k0}
Inhomogeneous parts of equations \eqref{Ameq0}-\eqref{cmeq0}:
\begin{equation}
\begin{split}
J_A=- \frac{\mu^2}{12x^{11}} (82707 Q^3 \mu-84196 Q^2 x \mu+32154 Q x^2 \mu+(912 Q^2-4320 \mu) x^3-160 Q x^4)
\end{split}
\eqlabel{sourceA0}
\end{equation}
\begin{equation}
\begin{split}
J_B=&\frac{\mu^2}{8x^{11}  (-x \mu+Q \mu+x^3)} \biggl(
704 Q x^7+(-11520 \mu-2304 Q^2) x^6+83536 Q \mu x^5\\
&-2232 \mu (96 Q^2-5 \mu) x^4+8 Q \mu (26678 Q^2-11669 \mu) x^3+ 294968 Q^2 \mu^2 x^2\\
&-423932 Q^3 \mu^2 x+ 210625 Q^4 \mu^2\biggr)
\end{split}
\eqlabel{sourceB0}
\end{equation}
\begin{equation}
\begin{split}
J_{\nu}=&\frac{\mu^2}{320 x^{11} (-x \mu+Q \mu+x^3)} \biggl(
47488 Q x^7-160704 Q^2 x^6-132544 Q \mu x^5+ 8 \mu (3915 \mu\\
&+70168 Q^2) x^4-48 Q \mu (10027 Q^2+3274 \mu) x^3+292748 Q^2 \mu^2 x^2-349232 Q^3 \mu^2 x\\
&+225375 Q^4 \mu^2\biggr)
\end{split}
\eqlabel{sourcenu0}
\end{equation}
\begin{equation}
\begin{split}
J_{a}=&\frac{\mu}{2 x^{11}} \biggl(-576 Q x^6+3840 \mu x^5-16720 Q \mu x^4+24 \mu (-210 \mu+1061 Q^2) x^3\\
&+ 26726 \mu^2 Q x^2-53200 Q^2 \mu^2 x+31473 Q^3 \mu^2\biggr)
\end{split}
\eqlabel{sourcena0}
\end{equation}
\begin{equation}
\begin{split}
J_{const}=&-\frac{\mu^2}{24 x^{10}} \biggl(1696 Q x^7-3968 Q^2 x^6-5656 Q \mu x^5+40 \mu (81 \mu+508 Q^2) x^4 \\
&-4 Q \mu (5211 Q^2+5110 \mu) x^3+53844 Q^2 \mu^2 x^2-69972 Q^3 \mu^2 x+37841 Q^4 \mu^2\biggr)
\end{split}
\eqlabel{sourcec0}
\end{equation}

Inhomogeneous part of equation \eqref{final0}:
\begin{equation}
\begin{split}
\cali_{\nu}=&-\frac{\mu^2}{320 x^{11} (-x \mu+Q \mu+x^3)}\biggl( -33920 Q x^7+ 122816 Q^2 x^6+ 118016 Q \mu x^5 \\
&-8 \mu (63224 Q^2+675 \mu) x^4+16 Q \mu (28147 Q^2-2078 \mu) x^3+260180 Q^2 \mu^2 x^2\\
&-423344 Q^3 \mu^2 x+ 189257 Q^4 \mu^2\biggr)
\end{split}
\eqlabel{io}
\end{equation}

Inhomogeneous part of equation \eqref{dil0}:
\begin{equation}
\begin{split}
J_{\phi}=&-\frac{9 \mu^2}{16 x^{11} (-\mu x+x^3+Q \mu)} \biggl( 1059 Q^4 \mu^2-1888 Q^3 \mu^2 x+1532 Q^2 \mu^2 x^2+(-192 Q^3 \mu\\
&-640 Q \mu^2) x^3+(320 \mu Q^2+120 \mu^2) x^4-160 Q \mu x^5+64 Q^2 x^6\biggr)
\end{split}
\eqlabel{jphi0}
\end{equation}

\section{Warp factors of the $\a'$-corrected charged $\k=0$ black hole to leading order in $Q$}\label{koq}
All expressions below are to order $\calo\left(Q^2\right)$ except for the gauge potential $a(x)$ which is evaluated 
to order $\calo\left(Q\right)$ .

Location of the horizon:
\begin{equation}
\xp=\sqrt{\mu}-\frac 12 Q
\eqlabel{eqw1}
\end{equation}

Metric warp factors:
\begin{equation}
A= -\frac{60\mu^3}{7x^6}+\left(-\frac {4\mu^2}{9x^5}+\frac{5359\mu^3}{112x^7}\right)\ Q
\eqlabel{eqAw1}
\end{equation}
\begin{equation}
\begin{split}
B=&\frac{15\mu}{14x^6} \biggl(37 \mu^2-19 x^2 \mu-19 x^4\biggr)-\frac{Q}{504(\mu-x^2) x^7} 
\biggl(32749 \sqrt{\mu} x^7+103971 \mu^4\\
&-136900 \mu^3 x^2+180 \mu^2 x^4\biggr) 
\end{split}
\eqlabel{eqBw1}
\end{equation}
\begin{equation}
\begin{split}
\nu=&\frac{15\mu^2(\mu+x^2)}{32x^6}+Q\biggl(-\frac{113241(2 x^2-\mu)}{5600\mu^{3/2}} \left(
\ln\left(1-\frac{\mu}{x^2}\right)+2 {\rm arctanh} \frac {\sqrt{\mu}}{x}\right) + \frac{9857\mu^3}{3920x^7}
\\
&-\frac{1144277 \mu^2}{98000x^5}+\frac{37747\mu}{14000x^3}-\frac{37747}{2800x}-\frac{113241}{2800\sqrt{\mu}}+\frac{113241x}{1400\mu}\biggr)
\end{split}
\eqlabel{eqnuw1}
\end{equation}

Gauge potential:
\begin{equation}
\begin{split}
a=&-\frac{193\mu^2}{2x^4}+\frac{835\mu^3}{14x^6}
\end{split}
\eqlabel{eqaw1}
\end{equation}

\end{document}